\documentclass[aps,prb,superscriptaddress,draft,showpacs,intlimits,amsmath,amssymb,floats,floatfix,twocolumn]{revtex4}
\usepackage[final]{graphicx}
\usepackage{epsfig}
\usepackage{xspace}

\newcommand{\LSCO}{\mbox{La$_\mathrm{2-x}$Sr$_\mathrm{x}$CuO$_4$}\xspace}
\newcommand{\TlBaCuO}{\mbox{Tl$_2$Ba$_2$CuO$_{6+\delta}$}\xspace}
\newcommand{\Tc}{\mbox{$T_\mathrm{c}$}\xspace}
\newcommand{\Aoneg}{\mbox{A$_\mathrm{1g}$}\xspace}
\newcommand{\Atwog}{\mbox{A$_\mathrm{2g}$}\xspace}
\newcommand{\Boneg}{\mbox{B$_\mathrm{1g}$}\xspace}
\newcommand{\Btwog}{\mbox{B$_\mathrm{2g}$}\xspace}

\begin{document}
\title{Electron interactions and charge ordering in La$_{2-x}$Sr$_x$CuO$_4$}
\author{B. Muschler}
\email{bernhard.muschler@wmi.badw.de}
\affiliation{Walther Meissner Institut, Bayerische Akademie der
Wissenschaften, 85748 Garching, Germany}
\author{W. Prestel}
\affiliation{Walther Meissner Institut, Bayerische Akademie der
Wissenschaften, 85748 Garching, Germany}
\author{L. Tassini}
\affiliation{Walther Meissner Institut, Bayerische Akademie der
Wissenschaften, 85748 Garching, Germany}
\author{R. Hackl}
\affiliation{Walther Meissner Institut, Bayerische Akademie der
Wissenschaften, 85748 Garching, Germany}
\author{M. Lambacher}
\affiliation{Walther Meissner Institut, Bayerische Akademie der
Wissenschaften, 85748 Garching, Germany}
\author{A. Erb}
\affiliation{Walther Meissner Institut, Bayerische Akademie der
Wissenschaften, 85748 Garching, Germany}
\author{Seiki Komiya}
\affiliation{Central Research Institute of the Electric Power 
Industry, Komae, Tokyo 201-8511, Japan}
\author{Yoichi Ando}
\affiliation{Institute of Scientific and Industrial Research,
Osaka University, Ibaraki, Osaka 567-0047, Japan}
\author{D.~C. Peets}
\altaffiliation[present address: ]{Department of Physics, 
Graduate School of Science, Kyoto University, Kyoto 606-8502, Japan}
\affiliation{Department of Physics and Astronomy, University 
of British Columbia, Vancouver V6T 1Z4, Canada}
\author{W.~N. Hardy}
\affiliation{Department of Physics and Astronomy, University 
of British Columbia, Vancouver V6T 1Z4, Canada}
\affiliation{Canadian Institute for Advanced Research, Toronto
 M5G 1Z8, Canada}
 \author{Ruixing Liang}
\affiliation{Department of Physics and Astronomy, University 
of British Columbia, Vancouver V6T 1Z4, Canada}
\affiliation{Canadian Institute for Advanced Research, Toronto
 M5G 1Z8, Canada}
 \author{D.~A. Bonn}
\affiliation{Department of Physics and Astronomy, University 
of British Columbia, Vancouver V6T 1Z4, Canada}
\affiliation{Canadian Institute for Advanced Research, Toronto
 M5G 1Z8, Canada}
\begin{abstract}
We present results of inelastic light scattering experiments on single-crystalline \LSCO in the doping range $0.00 \le x=p \le 0.30$ and\linebreak \TlBaCuO at $p=0.20$ and $p=0.24$.
The main emphasis is placed on the response of electronic excitations in the antiferromagnetic phase, in the pseudogap range, in the superconducting state, and in the essentially normal metallic state at $x \ge 0.26$, where no superconductivity could be observed.
In most of the cases we compare \Boneg and \Btwog spectra which project out electronic properties close to $(\pi,0)$ and $(\pi/2, \pi/2)$, respectively.
In the channel of electron-hole excitations we find universal behavior in \Btwog symmetry as long as the material exhibits superconductivity at low temperature.
In contrast, there is a strong doping dependence in \Boneg symmetry:
(i) In the doping range $0.20 \le p \le 0.25$ we observe rapid changes of shape and temperature dependence of the spectra.
(ii) In \LSCO new structures appear for $x < 0.13$ which are superposed on the electron-hole continuum.
The temperature dependence as well as model calculations support an interpretation in terms of charge-ordering fluctuations. For $x \le 0.05$ the response from fluctuations disappears at \Boneg and appears at \Btwog symmetry in full agreement with the orientation change of stripes found by neutron scattering.
While, with a grain of salt, the particle-hole continuum is universal for all cuprates the response from fluctuating charge order in the range $0.05 \le p < 0.16$ is so far found only in \LSCO.
We conclude that \LSCO is close to static charge order and, for this reason, may have a suppressed \Tc.
\end{abstract}
\maketitle
\section{Introduction}
\label{sec:intro}
The cuprates are fascinating materials not only for their superconducting transition temperatures $T_c$ in the 100\,K range \cite{Bednorz1986_ZPB64_189} but also for the wealth of new physics they exhibit.
The electron-phonon coupling mechanism which was believed to be almost universal (except for a few heavy fermion compounds and $^3$He) had to be supplemented if not abandoned.
This became experimentally evident when the gap was found to have $d$-wave \cite{Tsuei2000_RMP72_969} rather than $s$-wave symmetry as in conventional superconductors with reasonably high transition temperatures.
Another hallmark was the observation of electron dynamics with a strong anisotropy in the CuO$_2$ plane \cite{Damascelli2003_RMP75_473}.
A reverberation of the unconventional carrier dynamics was observed as early as 1989 in NMR experiments when studying the Knight shift \cite{Alloul1989_PRL63_1700}.
The unexpected decrease of the spin-lattice relaxation $1/T_1T$ towards lower temperature indicated a (pseudo)gap in the electron spectrum preceding the superconducting gap \cite{Norman2005_AP54_715}. The origin of the pseudogap and of superconductivity among several other instabilities is still elusive in spite of enormous progress in the experimental understanding.

Electronic Raman scattering (ERS) turned out to be instrumental to analyze properties of the cuprates \cite{Devereaux2007_RMP79_175}.
While gap anisotropies in the superconducting state were found early \cite{Hackl1988_PRB38_7133,Cooper1988_PRB38_11934}, the notion of gap nodes was indirect and qualitative at first \cite{Hackl1988_PRB38_7133,Monien1989_PRL63_911}.
In the normal state, carrier anisotropies in the $ab$-plane were experimentally obvious \cite{Staufer1990_SSC75_975,Slakey1991_PRB43_3764} but a quantitative understanding was possible only after 1994 when the relationship between scattering symmetries and different regions in momentum space was explained \cite{Devereaux1994_PRL72_396}.
It is in fact these selection rules which make ERS truly complementary to optical (IR) spectroscopy in which a similar type of two-particle response is measured with, however, some resolution in momentum space (due to the fluctuation nature of light scattering in charged systems). \cite{Prestel2010}.

The selection rules were crucial to pin down the interpretation of novel excitations superposed on the particle-hole continuum in underdoped \LSCO \cite{Venturini2002_P66_060502,Tassini2005_PRL95_117002}.
The low-energy peaks cropping up at low temperature even in the presence of long range antiferromagnetic order whenever the doping $p$ was non zero \cite{Tassini2008_PRB78_020511} could be assigned to charge-ordering fluctuations \cite{Caprara2005_PRL95_117004}.
In \LSCO the ``Aslamazov-Larkin'' peaks appear in \Boneg and \Btwog symmetry above and below $p=0.05$, respectively, compatible with the rotation of the dynamic superstructure or stripes observed by neutron scattering \cite{Fujita2002_PRB65_064505}.
This allows us to track ordering phenomena with short correlation lengths in the cuprates with Raman scattering.
Stripes seem to be omnipresent \cite{Kivelson2003_RMP75_1201} and may be of importance for both the pseudogap and superconductivity.
The well defined symmetry dependence of the response in \LSCO highlights the orientational order (nematicity).

In this publication we explore the generic properties and the interrelation of ordering phenomena in the cuprates by studying a very wide doping and temperature range.
We present new data on \LSCO (LSCO) and \TlBaCuO (Tl2201) and put them into context with existing work.
In Section~\ref{sec:2} we review some basic aspects of Raman scattering and show how annealing and aging affect the material properties and the Raman spectra.
Results on LSCO and Tl2201 are presented in section~\ref{sec:3}.
In section~\ref{sec:4} we discuss those properties of the cuprates which we consider generic or even universal.
We focus on spin and charge dynamics, fluctuation phenomena and finally the phase diagram with its crossover lines.

\section{Experiment and Samples}
\label{sec:2}

\subsection{Principles of electronic Raman scattering and selection rules}
\label{sec:20}
In a Raman experiment a photon flux density $I_0$ impinges on a sample and a number of photons per unit time $\dot{N}_{i,s}$ is scattered inelastically into an energy interval $\Delta\omega_s$ and a solid angle $\Delta\Omega$.
$\dot{N}_{i,s}$ corresponds to a set of incoming and outgoing photon polarizations, ${\bf e}_{i}$, ${\bf e}_{s}$.
From linear combinations of the measured spectra, pure symmetries $\mu$ can be derived. $\dot{N}_{i,s}$ can be written as
\begin{equation}
  \dot{N}_{i,s} = rI_0\frac{d^2\sigma_{i,s}}{d\Omega d\omega_s}T(\omega_s)\Delta \omega_s\Delta \Omega
  \label{eq:counts}
\end{equation}
where $\sigma_{i,s}$ is the cross section inside the material for certain polarizations, $r$ absorbs experimental factors such as losses at the sample surface or scattering volume and $T(\omega_s)$ is the (monochromatic) transmission coefficient of the complete set-up including the spectrometer and the CCD detector.
The spectra we show are generally response functions for a symmetry $\mu$, $R_{\mu}\chi^{\prime\prime}_{\mu}$, in units of counts per second and mW absorbed laser power [cts/(mW s)].
Among other factors $R_{\mu}$ includes matrix element effects since the response function $\chi_{\mu}({\bf q}, \Omega, T)$ is almost always determined in the non-resonant effective mass approximation \cite{Devereaux2007_RMP79_175}.
The connection between the internal cross section $\sigma_{\mu}$ for a given symmetry and the response functions $R_{\mu}\chi^{\prime\prime}_{\mu}$ is given by the fluctuation-dissipation theorem and can be expressed as
\begin{equation}
  \frac{d^2\sigma_{\mu}}{d\Omega d\omega_s} = \hbar r_0^2\frac{\omega_s}{\omega_i}\frac{1}{\pi}\{1+n(\Omega, T)\}R_{\mu}\chi^{\prime \prime}_{\mu}({\bf q} \rightarrow 0, \Omega, T).
  \label{eq:sigma}
\end{equation}
with $\Omega=\omega_i-\omega_s$ the excitation energy (Raman shift), $n(\Omega, T)$ the Bose-Einstein occupation number and $r_0=e^2/(4\pi\varepsilon_0 mc^2)$ the ``classical'' electron radius, hence Eq.~(\ref{eq:sigma}) describes the cross section per electron.
For the analyses presented below we do not need absolute cross sections (which could be calculated with some effort).
Independent of that, all spectra have the same scale and can be compared quantitatively to all of our previous results.
In particular, the measured count rates are normalized to a resolution of $\Delta \omega_s = 10$\,cm$^{-1}$ at 458\,nm and divided by $E(\omega_s)(\omega_s/\omega_i)\{1+n(\Omega, T)\}$ with $E(\omega_s)=T(\omega_s)\Delta \omega_s(\omega_s)$ the spectral sensitivity of the setup (see next paragraph).

Electronic Raman spectra (from particle-hole excitations) of symmetry $\mu$ project out specific regions in the Brillouin zone (BZ) \cite{Devereaux1994_PRL72_396}. In most of the cases here we show results in \Boneg and \Btwog symmetry where carriers with momenta along the principal axes and in the centers of the quadrants of the BZ, respectively, are emphasized.
The squares of the vertices are shown in the insets of Figs.~\ref{fig:cleavedpolished026} (a) and (b).
%%%%%%%%%%%%%%%%%%%%%%%%%%%%%%%%%%%%%%%%%%%%%%%%%%%%%%%%%%%%%%%%%%%%%%%%%%%%%%%%%%%%%%%%%%%%%%%%%%%%%%%%%%%%%%
\subsection{Raman experiment}
\label{subsec:21}
The samples were mounted on the cold finger of a He-flow cryostat in a vacuum of typically $5\times 10^{-7}$\,mbar.
All spectra were measured with an Ar$^{+}$ laser emitting in the range 458-514\,nm.
The spectra of LSCO were generally measured with an exciting wavelength of 458\,nm.
The data on Tl2201 were collected with the line at 514\,nm in order to reduce phonon resonances.
All light polarizations were controlled from outside the sample in such a way that the polarization state inside the sample had the desired orientation.
For the  laser light ($\omega_i$) incident at an angle of typically 66$^{\circ}$ to the surface normal this can be achieved by compensating for the phase shifts introduced at the sample surface.
The accuracy of the polarization state in the sample was always better than 99\,\%.
Since the scattered light ($\omega_s$) is collected along the surface normal its polarization state can be selected trivially.
The scattered light was analyzed in most of the cases with a subtractively coupled Jarrell-Ash 25-100 scanning spectrometer.
The image of the laser focus on the entrance slit was smaller than the slit width at all energy shifts, as verified with a CCD (charge coupled device) camera (see also \cite{Reznik1993_PRB48_7624}), to avoid losses due to the chromatic aberration of the collection optics.

In the energy range studied, changes in the optical constants are too small to influence the spectral shape significantly.
The efficiency $E(\omega_s)$ of the entire setup was determined with a standard lamp.
Using white light for the calibration determines both the pure monochromatic transmission $T(\omega_s)$ and the energy dependent variation of the spectral resolution $\Delta\omega_s(\omega_s)$ at constant slit width.
The transmission $T(\omega_s)$ can also be measured directly with monochromatic light while the resolution can be calculated to an accuracy of a few percent if the geometry of the spectrometer is known.
Therefore the efficiency $E(\omega_s) = T(\omega_s)\Delta\omega_s(\omega_s)$ can be determined in two completely independent ways.

\begin{figure}
  \centering
  \includegraphics{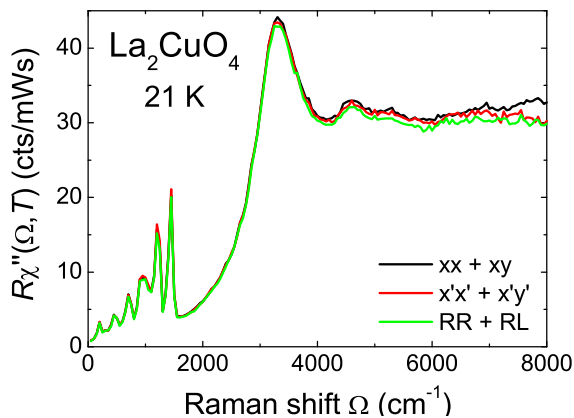}
  \caption[]{Sums of (Bose corrected) spectra at different polarization configurations, each projecting all four in-plane symmetries. The three sets agree to within statistical error up to 4000\,cm$^{-1}$. Above this energy the discrepancies are on the order of 10\% or less.
  } \label{fig:LCO_consist}
\end{figure}

With the optical constants determined accurately, the main polarizations in the $ab$-plane (see Table~\ref{tab:pol}) can be measured with high accuracy and can be checked for internal consistency as shown in Fig.~\ref{fig:LCO_consist} before the four symmetries $\mu=A_{1g}$, \Atwog, \Boneg, and \Btwog are determined. \Aoneg, for instance, is obtained as
\begin{equation}
A_{1g}=\frac{1}{3}\left[xx+x'x'+RR-\frac{1}{2}(xy+x'y'+RL)\right]
\label{eq:sym}
\end{equation}
where the light is always propagating along (or anti-parallel to) the $c$-axis and the polarizations are in the $ab$-plane with $x=[100]$, $x'=1/\sqrt{2}[110]$, $R=1/\sqrt{2}[1i0]$ etc.
%%%%%%%%%%%%%%%%%%%%%%%%%%%%%%%%%%%%%%%%%%%%%%%%%%%%%%%%%%%%%%%%
\begin{center}
\begin{table}
\caption[]{Symmetries $\mu$ accessible at the principal polarization directions in the basal plane of the tetragonal $D_{4h}$ point group. We use short-hand Porto notation with the first and the second symbol representing incoming (i) and outgoing (s) photon polarizations, respectively, since photons, to a good approximation, travel parallel and anti-parallel to the crystallographic $c$-direction. The polarizations are given as $x=[100]$, $y=[010]$, $x'=1/\sqrt{2}[110]$, $y'=1/\sqrt{2}[1\bar{1}0]$, $R_i=L_s=1/\sqrt{2}[1i0]$, and $R_s=L_i=1/\sqrt{2}[1\bar{i}0]$. The counterintuitive chirality of the circularly polarized photons arises from the back-scattering configuration.
  }
\centering
\begin{tabular}{c c}
  \hline\noalign{\smallskip}
  polarization   & symmetries $\mu$\\
  \noalign{\smallskip}\hline\hline\noalign{\smallskip}
  $xx = yy$      & $A_{1g} + B_{1g}$\\
  $xy = yx$      & $B_{2g} + A_{2g}$\\
  $x'x' = y'y'$  & $A_{1g} + B_{2g}$\\
  $x'y' = y'x'$  & $B_{1g} + A_{2g}$\\
  $RR = LL$      & $A_{1g} + A_{2g}$\\
  $RL = LR$      & $B_{1g} + B_{2g}$\\
  \noalign{\smallskip}\hline
\end{tabular}
  \label{tab:pol}
\end{table}
\end{center}
%%%%%%%%%%%%%%%%%%%%%%%%%%%%%%%%%%%%%%%%%%%%%%%%%%%%%%%%%%%%%%%%%%%%%%%%
%%%%%%%%%%%%%%%%%%%%%%%%%%%%%%%%%%%%%%%%%%%%%%%%%%%%%%%%%%%%%%%%%%%
%%%%%%%%%%%%%%%%%%%%%%%%%%%%%%%%%%%%%%%%%%%%%%%%%%%%%%%%%%%%%%%%%%%%%%%%
%%%%%%%%%%%%%%%%%%%%%%%%%%%%%%%%%%%%%%%%%%%%%%%%%%%%%%%%%%%%%%%%%%%
\begin{figure}
  \centering
  \includegraphics{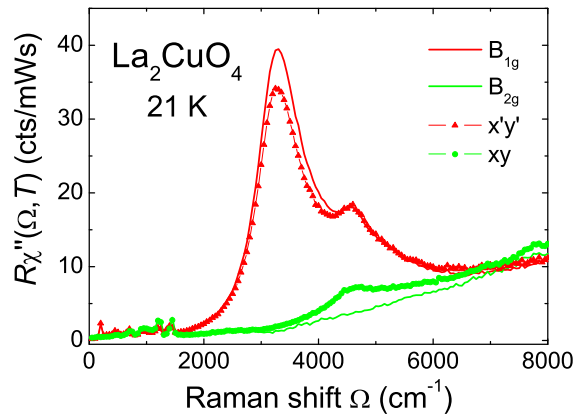}
  \caption[]{Comparison of spectra representing pure symmetries (\Boneg, \Btwog) multiplied by the factor $\omega_i/\omega_s$ (solid line) with directly measured spectra without multiplication by $\omega_i/\omega_s$ representing the sum of \Boneg+\Atwog and \Btwog+\Atwog (symbols).
  } \label{fig:sym_simplification}
\end{figure}
%%%%%%%%%%%%%%%%%%%%%%%%%%%%%%%%%%%%%%%%%%%%%%%%%%%%%%%%%%%%%%%%%%%%%%%%
We do not always present the response at pure symmetries, for the sake of saving measurement time, but the responses in $\Boneg + \Atwog$ and $\Btwog + \Atwog$ symmetry.
The analysis of many cuprates at various doping levels shows, however, that the error introduced by this simplification is small:
(i) At  energies $\Omega \leq 1000$\,cm$^{-1}$ the contributions from \Atwog excitations are generally an order of magnitude smaller than those at the other symmetries (Fig.~\ref{fig:LCO_puresym}).
(ii) For a freak of nature, in practically all \Boneg and \Btwog spectra the effect of \Atwog and the division by $(\omega_s/\omega_i)$ compensate each other to within $\pm 20$\% up to 8000\,cm$^{-1}$ Raman shift (see Fig.~\ref{fig:sym_simplification}).
(iii) Disorder in the sample can enhance the high energy cross section enormously (see below), and the inaccuracies due to the \Atwog contributions become significant only for extremely clean and ordered samples such as YBa$_{2}$Cu$_3$O$_{\mathrm{6+x}}$ prepared in BaZrO$_3$ crucibles \cite{Erb1995_PC245_245} or latest generation Bi$_{2}$Sr$_2$CaCu$_2$O$_{\mathrm{{8+\delta}}}$ \cite{Ando1995_PRL75_4662,Ando1996_PRL77_2065,Ando1997_PRB56_8530,Eisaki2004_PRB69_064512} (also see below).

\subsection{Samples}
\label{subsec:samples}
The samples for which we show results here are compiled in Table~\ref{tab:samples}.
All \LSCO ($0 \le x \le 0.30$), and ${\rm Bi_2 Sr_2 Ca Cu_2 O_{8+\delta}}$ samples were prepared in  image furnaces via the travelling solvent floating zone (TSFZ) technique. The \TlBaCuO crystals were prepared in crucibles \cite{Peets2010_JCG312_344}.
Generally, the samples were post-annealed (see table \ref{tab:samples}).
In some cases we also measured as grown samples for monitoring the changes via Raman spectroscopy.
The N\'eel temperatures $T_{\mathrm{N}}$ of antiferromagnetic (AF) samples and some of the superconducting transition temperatures \Tc were measured with the DC-SQUID.
In most of the cases \Tc was determined via the non-linear susceptibility.
%%%%%%%%%%%%%%%%%%%%%%%%%%%%%%%%%%%%%%%%%%%%%%%%%%%%%%%%%%%%%%%%%%%
\begin{figure*}
  \centering
  \includegraphics{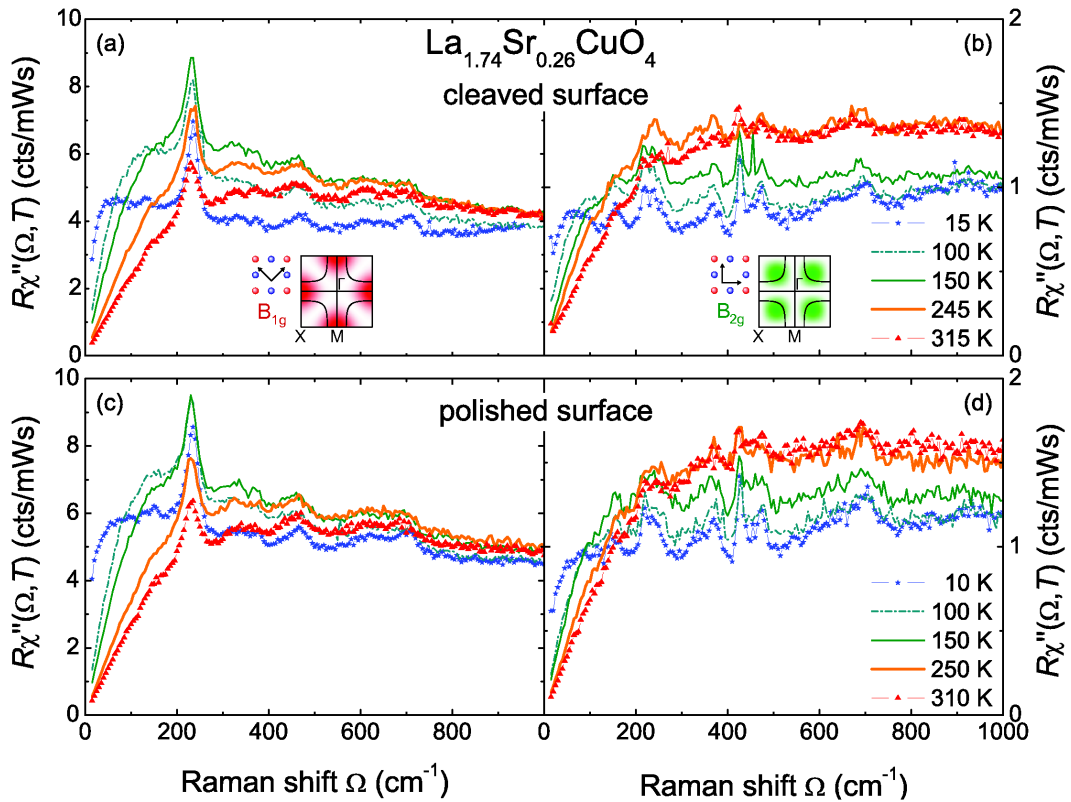}
  \caption[]{Comparison of the Raman response of \mbox{La$_{1.74}$Sr$_{0.26}$CuO$_4$} in \Boneg [(a),(c)] and \Btwog [(b),(d)] symmetry taken on cleaved [(a),(b)] and polished surfaces [(c),(d)]. The data do not show significant differences in the spectral shape nor in the temperature dependence. The insets represent the first BZ with the sensitive region for each symmetry with the corresponding light polarizations with respect to the oxygen (blue) and copper atoms (red).
  } \label{fig:cleavedpolished026}
\end{figure*}
%%%%%%%%%%%%%%%%%%%%%%%%%%%%%%%%%%%%%%%%%%%%%%%%%%%%%%%%%%%%%%%%%%%%%%%%
The spectra on LSCO were usually measured on polished surfaces.
The samples were polished in the crystal laboratory of the Technical University Munich.
For best results, diamond paste with a grain size of 0.3\,$\mu$m was used which we found yield better results than chemical etching.
For LSCO-26 we compared the results from polished and cleaved surfaces up to an energy transfer of 1000\,cm$^{-1}$ and did not find significant differences (see Fig.~\ref{fig:cleavedpolished026}).

%%%%%%%%%%%%%%%%%%%%%%%%%%%%%%%%%%%%%%%%%%%%%%%%%%%%%%%%%%%%%%%%
\begin{center}
\begin{table*}
\caption[]{Complete list of samples studied. Some results from those at the bottom of the table have been published in various earlier papers \cite{Tassini2005_PRL95_117002,Tassini2008_PhdThesis,Hackl2005_45_227,Hackl2006_JPCS67_289}. Samples labeled with $a$ have been prepared by M. Lambacher and A. Erb (WMI Garching) \cite{Erb2010}, $b$ by Shimpei Ono, Seiki Komiya and Yoichi Ando (CRIEPI, Tokyo  and Osaka University), $c$ by D. Peets, Ruixing Liang and D. Bonn (Vancouver) \cite{Peets2010_JCG312_344}, and $d$ by N. Kikugawa and T. Fujita (Hiroshima and Tokyo). The transition temperatures were measured either resistively or via magnetometry or via the non-linear ac response. The \Tc of the 5\% sample is the onset point of the transition. $T_N$ of \LSCO at $x=0.02$, 0.04 and 0.05 was not measured. In the latter two cases $T_N=0$.
}
\centering
\begin{tabular}{c c c c c c c}
 \hline\noalign{\smallskip}
 sample & sample ID & doping & ~~~~~$T_c/T_N$ (K) & $\Delta T_c$ (K)&  comment & \\
 \noalign{\smallskip}\hline\hline\noalign{\smallskip}
 ${\rm La_{2} CuO_{4}}$                            & LCO-00     & 0.00 & 0/270 & - & as grown       & $a$ \\
 %\hline
 ${\rm La_{2} CuO_{4}}$                            & LCO-00-Ar  & 0.00 & 0/325 & - & Ar annealed    & $a$ \\
 %\hline
 ${\rm La_{1.96}Sr_{0.04}CuO_{4}}$                 & LSCO-04    & 0.04 & 0/-   & - & O$_2$ annealed & $a$ \\
 %\hline
 ${\rm La_{1.95}Sr_{0.05}CuO_{4}}$                 & LSCO-05    & 0.05 & 5/-   & 3 & O$_2$ annealed & $a$ \\
 %\hline%%%%%%%%%%%%%%%%%%%%%%%%%%%%%%%%%%%%%%%%%%%
 ${\rm La_{1.92}Sr_{0.08}CuO_{4}}$                 & LSCO-08    & 0.08 & 18    & 4 & O$_2$ annealed & $d$ \\
 %\hline%%%%%%%%%%%%%%%%%%%%%%%%%%%%%%%%%%%%%%%%%%%
 ${\rm La_{1.85}Sr_{0.15}CuO_{4}}$                 & LSCO-15    & 0.15 & 38    & 3 & O$_2$ annealed & $a$ \\
 %\hline%%%%%%%%%%%%%%%%%%%%%%%%%%%%%
 ${\rm La_{1.83}Sr_{0.17}CuO_{4}}$                 & LSCO-17    & 0.17 & 39    & 1 & O$_2$ annealed & $b$ \\
 %\hline%%%%%%%%%%%%%%%%%%%%%%%%%%%%%%%%%%%%%%%%%%%
 ${\rm La_{1.80}Sr_{0.20}CuO_{4}}$                 & LSCO-20    & 0.20 & 24    & 3 & O$_2$ annealed & $a$ \\
 %\hline
 ${\rm La_{1.75}Sr_{0.25}CuO_{4}}$                 & LSCO-25    & 0.25 & 12    & 3 & O$_2$ annealed & $a$ \\
 %\hline
 ${\rm La_{1.70}Sr_{0.30}CuO_{4}}$                 & LSCO-30    & 0.30 & 0     & - & as grown       & $a$ \\
 %\hline
 ${\rm La_{1.70}Sr_{0.30}CuO_{4}}$                 & LSCO-30-O2 & 0.30 & 0     & - & 100\,bar O$_2$ & $a$ \\
 \noalign{\smallskip}\hline\noalign{\smallskip}%%%% Peets %%%%%%%%%%%%%%%%%%%%%%%%%%%%%%%%%%%%%%%%%%%

  \TlBaCuO                                         & Tl2201-20  & 0.20 & 78    & 5 & O$_2$ annealed & $c$ \\
 %\hline
  \TlBaCuO                                         & Tl2201-24  & 0.24 & 46    & 5 & O$_2$ annealed & $c$ \\

 \noalign{\smallskip}\hline\noalign{\smallskip}%%%%%%%%%%%%%%%%%%%%%%%%%%%%%%%%%%%%%%%%%%

  ${\rm Bi_2 Sr_2 Ca Cu_2 O_{8+\delta}}$           & Bi2212-16 & 0.16  & 94    & 2 & air annealed & $b$ \\

 \noalign{\smallskip}\hline\noalign{\smallskip}%%%%%%%%%%%%%%%%%%%%%%%%%%%%%%%%%

  ${\rm La_{1.98}Sr_{0.02}CuO_{4}}$                 & LSCO-02  & 0.02 & 0/-    & - &       as grown    & $d$ \\

  ${\rm La_{1.90}Sr_{0.10}CuO_{4}}$                 & LSCO-10  & 0.10 & 25     & 4 & O$_2$ annealed    & $d$ \\

  ${\rm La_{1.85}Sr_{0.15}CuO_{4}}$                 & LSCO-15  & 0.15 & 38     & 4 & O$_2$ annealed    & $d$ \\

  ${\rm La_{1.74}Sr_{0.26}CuO_{4}}$                 & LSCO-26  & 0.26 & 0     & - & O$_2$ annealed    & $d$ \\

  \noalign{\smallskip}\hline
\end{tabular}
  \label{tab:samples}
\end{table*}
\end{center}
%%%%%%%%%%%%%%%%%%%%%%%%%%%%%%%%%%%%%%%%%%%%%%%%%%%%%%%%%%%%%%%%%%%%%%%%
%%%%%%%%%%%%%%%%%%%%%%%%%%%%%%%%%%%%%%%%%%%%%%%%%%%%%%%%%%%%%%%%%%%%%%%%
\begin{figure}
  \centering
  \resizebox{0.9\columnwidth}{!}{\includegraphics{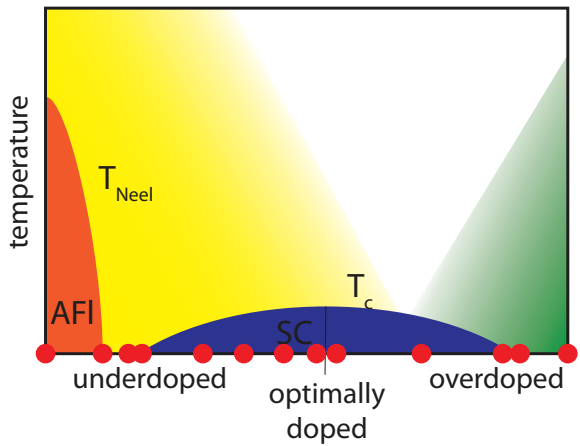}}
  \caption{
  Schematic phase diagram of hole-doped cuprates. The doping levels of the \LSCO crystals investigated in this study are indicated by dots.
  } \label{Fig:LSCO_doping}
\end{figure}
%%%%%%%%%%%%%%%%%%%%%%%%%%%%%%%%%%%%%%%%%%%%%%%%%%%%%%%%%%%%%%%%%%%%%%%%
\subsection{Annealing and aging}
\label{subsec:aging}
For all crystals the growth conditions depend crucially on the atmosphere, in particular on the oxygen partial pressure (see also \cite{Erb2010}).
Usually, the environment optimal for crystal growth does not return samples at the desired doping level or with the expected intrinsic properties, necessitating post-annealing.
%%%%%%%%%%%%%%%%%%%%%%%%%%%%%%%%%%%%%%%%%%%%%%%%%%%%%%%%%%%%%%%%%%%
\begin{figure*}
  \centering
  \includegraphics{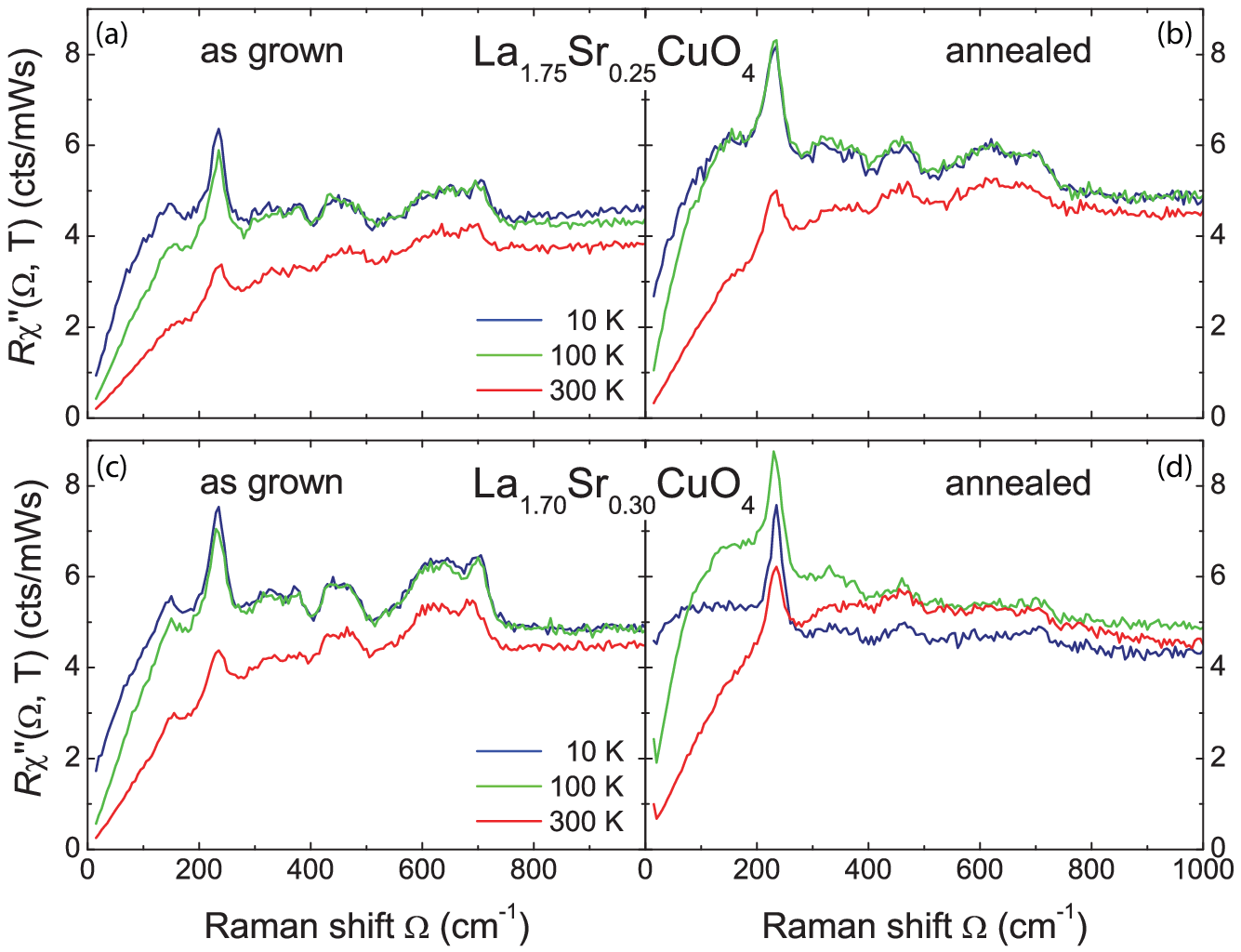}
  \caption[]{
  Effect of oxygen annealing on the Raman spectra in \Boneg symmetry of overdoped \LSCO at $x=0.25$ [(a),(b)] and $x=0.30$ [(c),(d)]. For both doping levels the spectral shape changes. Note the almost complete suppression of the modes between 600 and 800\,cm$^{-1}$ in high pressure annealed LSCO-30-O2. The high energy of the modes before annealing (c) indicates that mainly oxygen defect modes contribute. (d) Additionally the temperature dependence for $x=0.30$ becomes non-monotonic.}
  \label{fig:LSCO_anneal}
\end{figure*}
%%%%%%%%%%%%%%%%%%%%%%%%%%%%%%%%%%%%%%%%%%%%%%%%%%%%%%%%%%%%%%%%%%%%%%%%
For instance, LCO-00 had to be annealed in Ar for the N\'eel temperature $T_{\mathrm{N}}$ to reach 325\,K and an annealing in 1\,bar of O$_2$ improves the crystal quality for most Sr-containing samples \cite{Fujita__}.
For $x=0.25$ and $x=0.30$ on the overdoped side (see Table~\ref{tab:samples}), substantial differences in both the spectral shape and the temperature dependence of the electronic spectra in \Boneg symmetry were observed between as grown [Fig.~\ref{fig:LSCO_anneal}(a),(c)] and post-annealed samples [Fig.~\ref{fig:LSCO_anneal}(b),(d)].
In LSCO the only allowed mode in \Boneg symmetry is the out-of-phase vibration of the La atoms at 240\,cm$^{-1}$.
All the other bands at 150\,cm$^{-1}$, 400-500\,cm$^{-1}$, and 600-700\,cm$^{-1}$ are activated by lattice defects.
High pressure oxygen loading suppresses these bands substantially.
Apparently, they are activated by oxygen defects meaning that phonons with the proper symmetry are projected on the deficiency site \cite{Turrell1972__} and become Raman active.
In addition to the reduction of the defect bands by high-pressure oxygen loading the electronic continuum develops more structure:
(i) The initial slope of the spectra at 10\,K is much steeper for the annealed samples with $x=0.25$ and $x=0.30$ indicating a reduced elastic relaxation \cite{Devereaux2007_RMP79_175}.
(ii) Between 400 and 1000\,cm$^{-1}$ the continuum changes its shape from convex before to concave after annealing.
(iii) While the temperature dependence of the spectra of LSCO-25 changes only little after annealing at 1\,bar O$_2$, there is a significant change for LSCO-30 which was post-annealed at 100\,bar O$_2$.
Apparently, the oxygen deficiency at higher Sr concentrations is harder to eliminate.
For LSCO-30 the temperature dependence of the spectra becomes non-monotonic.
The spectrum at 100\,K exhibits a maximum similar to that of the sample with $x=0.26$ which will be discussed in Ref.~\cite{Prestel2010}. Note that there is hardly any change of the continuum below 100\,K in both as-grown samples.
We conclude that in the annealed crystals dynamic relaxation starts to dominate while in the as grown sample momentum scattering on defects dominates the relaxation at low temperature.
%%%%%%%%%%%%%%%%%%%%%%%%%%%%%%%%%%%%%%%%%%%%%%%%%%%%%%%%%%%%%%%%%%%%%%%%

\begin{figure}
  \centering
  \includegraphics{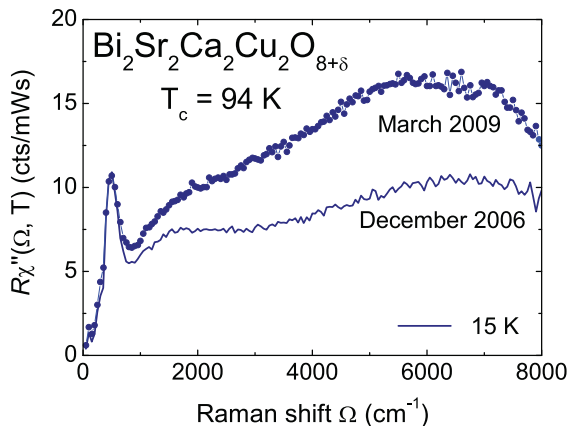}
  \caption{
  Effect of aging on Bi2212 at $x=0.16$. While the differences are weak below approximately 1000\,cm$^{-1}$ there is a substantial increase of intensity for higher energies.
  } \label{fig:Bi2212_aging}
\end{figure}

Finally, we found very strong changes of the high-energy continuum of Bi2212-16 as a function of sample age as shown in Fig.~\ref{fig:Bi2212_aging}.
At 8000\,cm$^{-1}$ the continuum is almost twice as intense for the aged sample as for the freshly prepared one.
We note that no change in \Tc, in $\Delta T_{\mathrm{c}}$, nor in the low energy part of the spectra could be observed.
It is known that Bi2212 is not completely stable and we speculate that out-of-plane defects trap charge and produce an enhanced level of luminescence.
No aging effects were observed for LSCO and ${\rm YBa_2Cu_3O_{\mathrm{6+x}}}$.
%%%%%%%%%%%%%%%%%%%%%%%%%%%%%%%%%%%%%%%%%%%%%%%%%%%%%%%%%%%%%%%%%%%
%%%%%%%%%%%%%%%%%%%%%%%%%%%%%%%%%%%%%%%%%%%%%%%%%%%%%%%%%%%%%%%%%%%%%%%%

This summary of effects of sample quality highlights that the Raman spectra are rather sensitive to disorder but can help to pinpoint the nature of the changes in some cases.
The defects appear to be away from the CuO$_2$ plane in Bi2212 while in overdoped LSCO the defects existing prior to annealing are located in the CuO$_2$ plane.
%%%%%%%%%%%%%%%%%%%%%%%%%%%%%%%%%%%%%%%%%%%%%%%%%%%%%%%%%%%%%%%%%%%%%%%%
\section{Results}
\label{sec:3}
We describe now the results on LSCO and Tl2201.
One purpose of the experiments on Tl2201 was to check whether or not the rather abrupt spectral changes in \Boneg symmetry in overdoped Bi2212 \cite{Venturini2002_PRL89_107003,Blanc:2009} are generic.
This issue cannot be settled straightforwardly with LSCO since the sample homogeneity in the range $0.19 \le x \le 0.25$ is reduced for thermodynamic reasons, and a transition from a hole- to an electron-like Fermi surface takes place \cite{Yoshida2006_PRB74_224510}.
However, in LSCO the entire doping range can be accessed in a single compound class and, bearing the sample quality issues in mind, the evolution of all excitations can be studied in detail.

\subsection{\LSCO}
\label{sec:31}

In this subsection all results on LSCO we have obtained over the years are compiled.
Those which are new can be identified in Table~\ref{tab:samples}.
We start with spectra at high energy transfers where spin and electron-hole excitations contribute.
Then the particle-hole dynamics in the range up to 1000~cm$^{-1}$ are presented and analyzed.
Finally, we compile the data on the fluctuation response obtained on the underdoped side which complete and supplement the study at $x=0.02$ and 0.10 \cite{Tassini2005_PRL95_117002}.
\subsubsection{Response at high energies}
\label{sec:311}

\begin{figure*}
  \centering
  \includegraphics{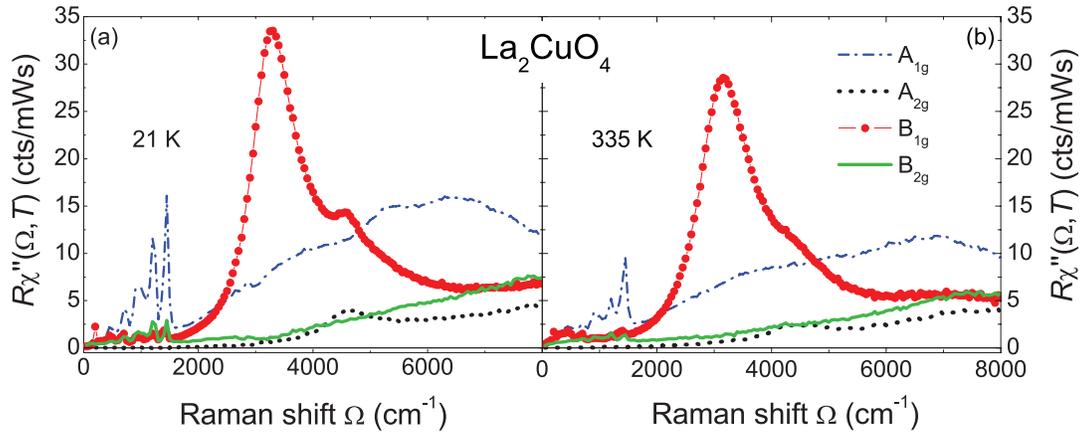}
  \caption[]{
  Raman response of antiferromagnetic La$_2$CuO$_4$ ($T_N=325$\,K) at 21 and 335\,K for all four symmetries observable in the $ab$-plane.
  } \label{fig:LCO_puresym}
\end{figure*}

In Fig.~\ref{fig:LCO_puresym} we plot the Raman response $R\chi^{\prime \prime}_{\mu}(\Omega, T)$ for two temperatures at all four symmetries of the tetragonal point group $D_{4h}$ which is by and large applicable for the high-energy part of the spectra.
Similar results have been published for Gd$_2$CuO$_4$ at room temperature \cite{Sulewski1991_PRL67_3864} and, for a smaller set of polarizations, for various other cuprates \cite{Reznik1993_PRB48_7624,Sugai1988_PRB38_6436,Blumberg1994_PRB49_13295,Blumberg1996_PRB53_11930,Sugai2003_PRB68_184504,Gozar2005__755}.
Due to the orthorhombic distortion below about 500\,K, a small $a-b$ anisotropy is also found at high energies - in addition to a strong one in the energy range where phonons occur \cite{Gozar2005__755}.
Up to 4000\,cm$^{-1}$, the data agree well with those on \mbox{Gd$_2$CuO$_4$} \cite{Sulewski1991_PRL67_3864}.
Similarly, qualitative agreement is also found with results on other cuprates \cite{Reznik1993_PRB48_7624,Blumberg1994_PRB49_13295,Blumberg1996_PRB53_11930,Gozar2005__755}.
The structures at low temperature between 1000 and 2000\,cm$^{-1}$ are due to two-phonon processes \cite{Sugai1989_PRB39_4306} which are observed only in the undoped compound and become weaker at high temperature [Fig.~\ref{fig:LCO_puresym}(b)].
At low temperatures  new structures appear in \Aoneg and \Boneg symmetry at 5200 and 4500\,cm$^{-1}$, respectively, and the \Atwog peak at 4600\,cm$^{-1}$ becomes sharper.
This \Boneg peak is most pronounced at low temperatures and doping levels and becomes broader for higher temperatures.
The results are shown for comparison and will be discussed in more detail elsewhere.

%%%%%%%%%%%%%%%%%%%%%%%%%%%%%%%%%%%%%%%%%%%%%%%%%%%%%%%%%%%%%%%%%%%
%%%%%%%%%%%%%%%%%%%%%%%%%%%%%%%%%%%%%%%%%%%%%%%%%%%%%%%%%%%%%%%%%%%%%%%%

\begin{figure*}
  \centering
  \includegraphics{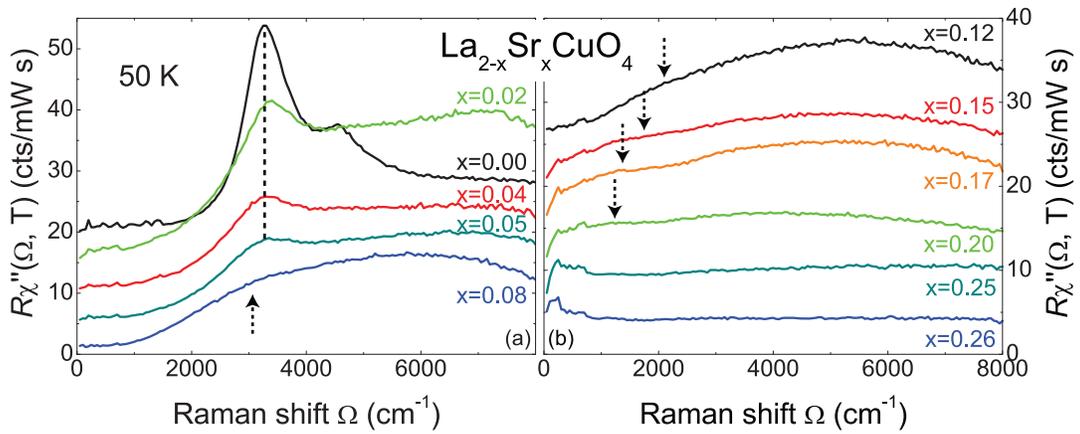}
  \caption[]{
  High-energy Raman response  at 50\,K of \LSCO in \Boneg symmetry for characteristic doping levels as indicated. The spectra of the samples at $x=0$ and 0.02 are taken at 90\,K. Successive data sets are offset vertically by 5 units each for clarity. The dashed vertical line in (a) marks the position of the two-magnon excitation for $x\leq 0.05$, the arrows indicate the position of the two-magnon excitation for higher doping levels. For $x\ge0.26$ no peak is observable.
  } \label{fig:LSCO_long}
\end{figure*}

At higher doping levels the response from spin excitations decreases rapidly \cite{Reznik1993_PRB48_7624,Blumberg1994_PRB49_13295,Blumberg1996_PRB53_11930,Sugai2003_PRB68_184504,Gozar2005__755} and can be observed for $x \geq 0.02$ in \Boneg symmetry only.
Therefore we show in Fig.~\ref{fig:LSCO_long} only \Boneg spectra at low temperature in the doping range $0 \le x \le 0.26$.
As a general trend, the entire spectra up to 8000\,cm$^{-1}$ decrease continuously in intensity towards higher doping.
At the highest doping levels the response above 2000\,cm$^{-1}$ is essentially flat in \Boneg symmetry.
The two-magnon peak becomes very weak above 5\% doping but may still be resolved in the range 1000 to 1500\,cm$^{-1}$ at 25\%.
The well-defined peak in \Boneg symmetry below 1000\,cm$^{-1}$ at $x=0.26$ is clearly resulting from electron-hole excitations as will be discussed briefly in subsection~\ref{sec:312} and in detail in the contribution by Prestel \textit{et al.} \cite{Prestel2010}.
%%%%%%%%%%%%%%%%%%%%%%%%%%%%%%%%%%%%%%%%%%%%%%%%%%%%%%%%%%%%%%%%%%%
%%%%%%%%%%%%%%%%%%%%%%%%%%%%%%%%%%%%%%%%%%%%%%%%%%%%%%%%%%%%%%%%%%%%%%%%
\subsubsection{Electron-hole excitations in the limit $\Omega \rightarrow 0$}
\label{sec:312}
\begin{figure*}
  \centering
  \includegraphics{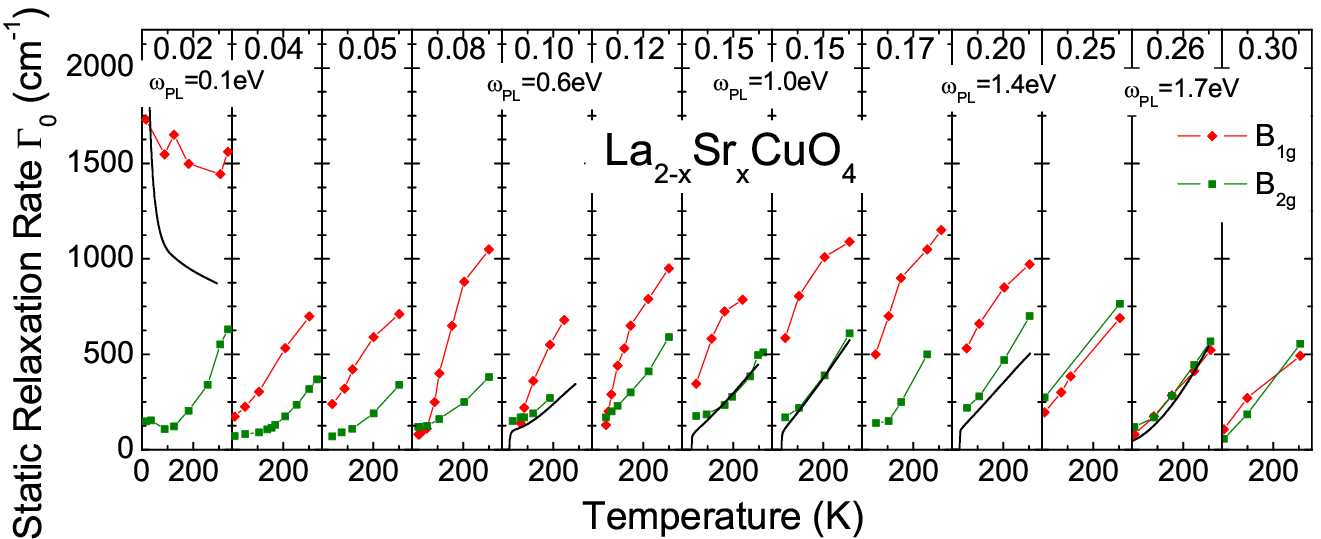}
  \caption[]{
  Raman resistivities of \LSCO at \Boneg and \Btwog symmetry at doping levels as indicated. The errors are of the size of the symbols. The solid lines are resisitivity data of the very same crystals as communicated by T. Fujita \cite{Fujita__} and A. Erb \cite{Erb2010}. The ``Raman resistivity'' is the inverse initial slope of the Raman response corresponding to $\Gamma_{0,\mu}(T) = \hbar/\tau_{\mu}$ where $\tau_{\mu}$ is a symmetry-dependent carrier relaxation time. This is related but not equal to $\tau_{opt}$ in the optical conductivity (for a discussion see ref.~\cite{Opel2000_PRB61_9752}).
  } \label{fig:LSCO_Gamma}
\end{figure*}

In this section we show only analyzed data.
The technique employed to derive ``Raman resistivities'', i.~e. the inverse slopes of the spectra, $\Gamma_{0,\mu}(T) = [\partial \chi^{\prime \prime}_{\mu}(\Omega,T)/\partial \Omega|_{\Omega \rightarrow 0}]^{-1}$, from the response is described in detail by Opel \textit{et al.} \cite{Opel2000_PRB61_9752}.
In Fig.~\ref{fig:LSCO_Gamma} all results we obtained for LSCO so far are compiled.
There are four major observations:
(i) For $x \ge 0.25$ the rates $\Gamma_{0,\mu}(T)$ for \Boneg and \Btwog symmetry are similar in both magnitude and temperature dependence indicating more or less isotropic metallic behavior.
$\Gamma_{0,B1g}(T)$ and $\Gamma_{0,B2g}(T)$ closely track the resistivity when converted via the Drude approximation $\Gamma_{0}(T)=1.08\,\omega_{pl}^{2}\,\rho(T)$, with the plasma frequency $\omega_{pl}$ in units of eV and the resistivity $\rho(T)$ in $\mu\Omega$cm.
(ii) Upon reducing the doping from 0.25 to 0.20 $\Gamma_{0,B1g}(T)$ and $\Gamma_{0,B2g}(T)$ become different.
(iii) For $0.05 \le x \le 0.20$ we find substantial differences in the magnitudes of $\Gamma_{0,B1g}(T)$ and $\Gamma_{0,B2g}(T)$ at room temperature. The temperature dependences are metallic in either case with $\partial\Gamma_{0,B1g}(T)/\partial T$ increasing towards underdoping.
Only $\Gamma_{0,B2g}(T)$ matches the resistivity.
(iv) For $x \le 0.02$, none of the Raman resistivites exhibits similarities with the temperature dependence of the resistivity.
$\Gamma_{0,B1g}(T)$ is much larger than $\Gamma_{0,B2g}(T)$ for all temperatures and weakly non-metallic.

As we will see in section~\ref{sec:32}, the crossover in the range $0.20 \le p=x \le 0.25$ is also present in Tl2201.
Due to the uniquely wide doping range accessible in LSCO the other relevant doping levels at approximately 5 and 25\% are also observable.
Are there additional features?
%%%%%%%%%%%%%%%%%%%%%%%%%%%%%%%%%%%%%%%%%%%%%%%%%%%%%%%%%%%%%%%%%%%
%%%%%%%%%%%%%%%%%%%%%%%%%%%%%%%%%%%%%%%%%%%%%%%%%%%%%%%%%%%%%%%%%%%%%%%%

\begin{figure*}
  \centering
  \includegraphics{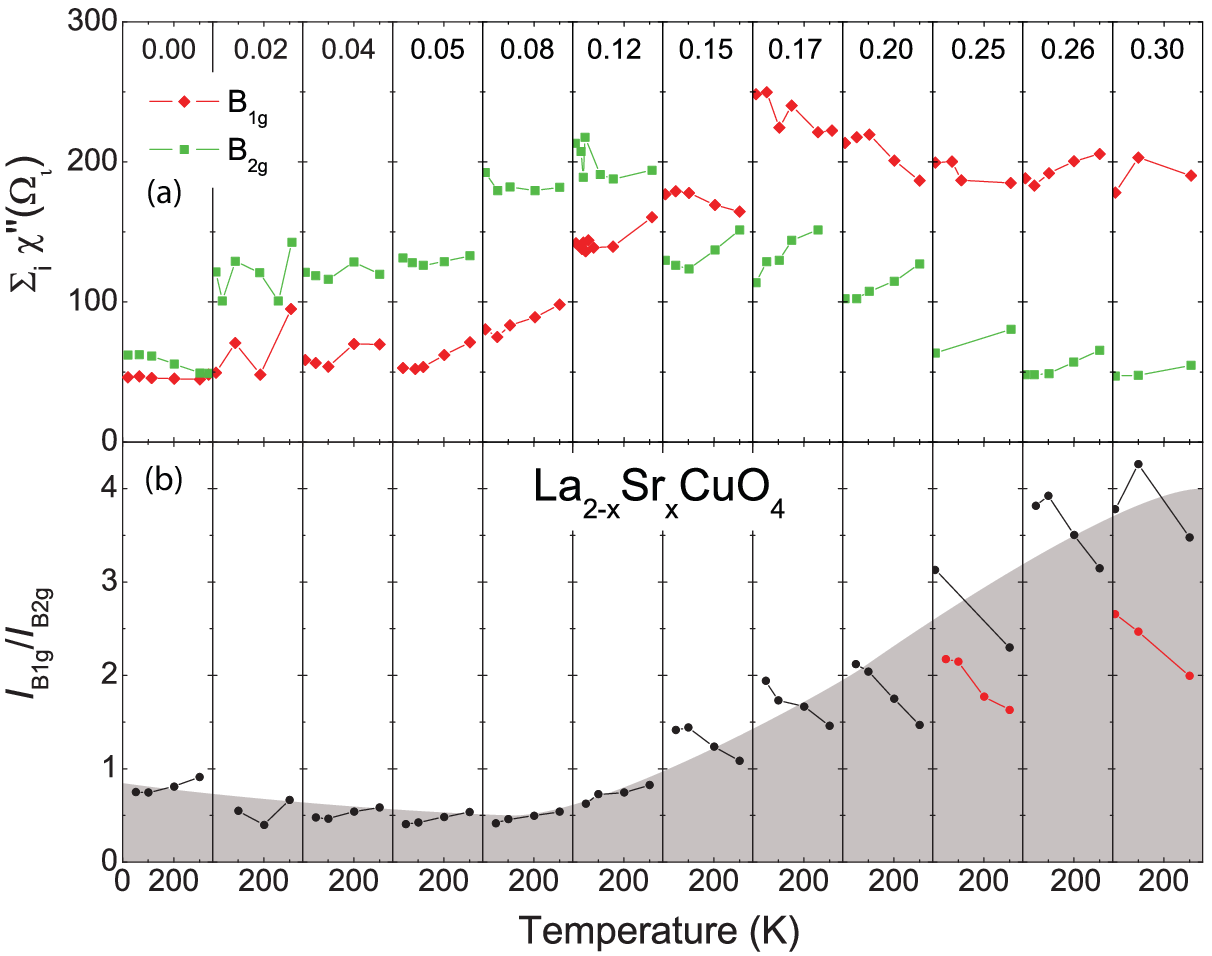}
  \caption[]{
  (a) Integrated intensities of the Raman spectra in \Boneg and \Btwog symmetry at doping levels as indicated. The integration range is limited to energies between 800 and 1000\,cm$^{-1}$, where the spectra are relatively temperature independent. (b) ratio of the integrated \Boneg and \Btwog Raman spectra, $I_{\mathrm{B1g}}/I_{\mathrm{B2g}}$. The red (gray) curves are the result of a analysis of the as-grown samples. The statistical errors are of the size of the symbols. The shaded area illustrates the gain in the ratio $I_{\mathrm{B1g}}/I_{\mathrm{B2g}}$ with doping.}
\label{fig:LSCO_ratio}
\end{figure*}

To answer this question we look at a different quantity, the integrated intensities of the spectra in \Boneg and \Btwog symmetry, $I_{\mathrm{B1g}}$ and $I_{\mathrm{B2g}}$, respectively, and the ratio of the intensities, $I_{\mathrm{B1g}}/I_{\mathrm{B2g}}$, integrated in the range from 800 up to 1000\,cm$^{-1}$, where the spectra are only weakly temperature dependent \cite{Opel2000_PRB61_9752}.
The results are compiled in Fig.~\ref{fig:LSCO_ratio}.
$I_{\mathrm{B1g}}$ exhibits little doping dependence for $x \leq 0.05$, increases for $0.08 \leq x \leq 0.17$, and becomes again doping independent for higher doping levels.
In contrast, $I_{\mathrm{B2g}}$ shows a maximum at approximately 12\% doping and decreases on either side.
The temperature dependences in the two symmetries are different as well:
While the temperature dependence of $I_{\mathrm{B2g}}$ does not change significantly with doping, $I_{\mathrm{B1g}}$ increases slightly with temperature for low doping, while on the overdoped side it decreases with increasing temperature.
In a tight binding scheme $I_{\mathrm{B1g}}/I_{\mathrm{B2g}} = (t/2t^{\prime})^2$ is expected to hold, with $t$ and $t^{\prime}$ the nearest and next-nearest neighbor hopping parameters, respectively \cite{Einzel1996_JRS27_307}.
This is the case at high doping levels (see Ref.~\cite{Prestel2010}) and should not change substantially for lower doping as pointed out in Refs.~\cite{Katsufuji1993_PRB48_16131,Naeini1999_PRB59_9642}.
In contrast to the expectation, not only does the ratio decrease by almost an order of magnitude, but the temperature derivative changes from negative to slightly positive in the range $0.12 < x < 0.15$.
Hence, there is another crossover close to optimal doping, which appeared also in the relaxation rates of Bi2212 \cite{Venturini2002_PRL89_107003}.
The question arises as to how these properties fit into a universal picture and what may explain the discrepancies.
%%%%%%%%%%%%%%%%%%%%%%%%%%%%%%%%%%%%%%%%%%%%%%%%%%%%%%%%%%%%%%%%%%%
%%%%%%%%%%%%%%%%%%%%%%%%%%%%%%%%%%%%%%%%%%%%%%%%%%%%%%%%%%%%%%%%%%%%%%%
\subsubsection{Density-wave fluctuations}
\label{sec:313}

LSCO opens a window towards an answer.
We hypothesize that compounds with a maximal transition temperature $T_{\mathrm{c}}^{\max}$ close to 100\,K show ``canonical'' carrier dynamics while in those with reduced $T_c^{\max}$ the generic properties may be masked.
Only for Bi2212 ($T_{\mathrm{c}}^{\max} \ge 94\,K$) have the carrier dynamics in the normal state been studied in detail \cite{Venturini2002_PRL89_107003}.
The body of data in Y123 ($T_{\mathrm{c}}^{\max}=93\,K$) is much smaller and the \Boneg electronic response cannot directly be accessed since the spectra of the CuO$_2$ planes are superposed by the strongly coupling buckling mode \cite{Opel1999_JLTP117_347,Opel:1999} and the response from the CuO chains.
Nevertheless, we find a clear dichotomy between the \Boneg and \Btwog carrier dynamics below $p \approx 0.20$ in that $\Gamma_{0,B2g}(T) \propto \rho(T)$ and $\Gamma_{0,B1g}(T)$ is approximately temperature independent \cite{Staufer1990_SSC75_975,Slakey1991_PRB43_3764,Venturini2002_PRL89_107003,Opel2000_PRB61_9752,Zhou1996_SSC99_669}.
In section~\ref{sec:32} we present new results on Tl2201 with $T_c^{\max} \ge 90\,K$ that fit into this framework.
In which sense is LSCO different?

\begin{figure}
  \centering
  \resizebox{0.86\columnwidth}{!}{\includegraphics{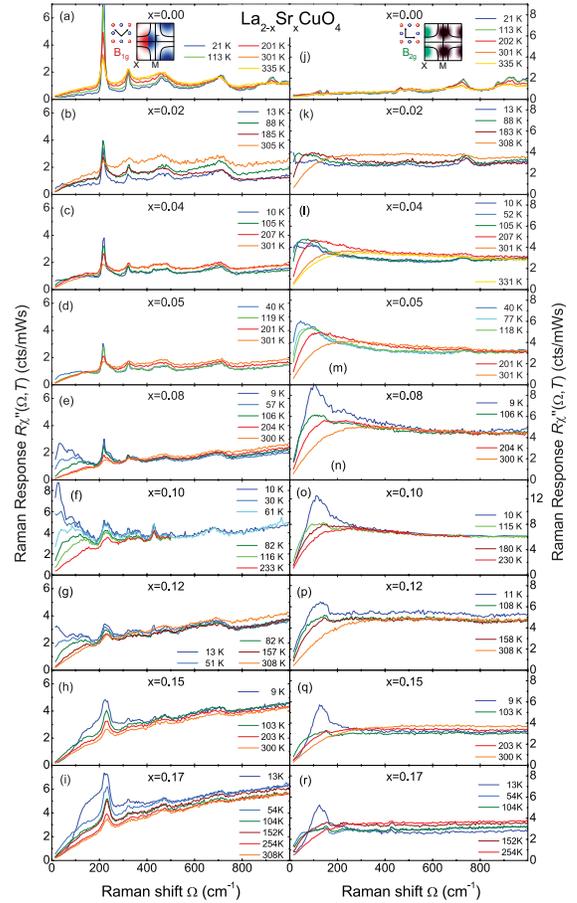}}
  \caption[]{
  Low energy Raman response of LSCO in the doping range $0.00 \le x \le 0.17$ in \Boneg and \Btwog symmetries at doping levels as indicated. The Aslamazov-Larkin type peaks flip symmetry at $x_{c}\approx0.05$ and are observable in \Boneg only for $0.08\leq x\leq 0.12$ [(e)-(g)] and in \Btwog symmetry for $0.02 \leq x \leq 0.5$ [(k)-(m)]. Note that the vertical scales change for different doping levels. Similar spectra were published by Sugai and coworkers \cite{Sugai2006_PRL96_137003}. However, there are important differences in the details. For instance, there is no fine structure in the \Btwog fluctuation peak at $x=0.05$ (m) which allows an interpretation in terms of multiple contributions. In contrast, the sample with the lowest quality ($x=0.08$) exhibits a rich substructure in \Boneg symmetry (e).
  }
\label{fig:LSCO_UD_series}
\end{figure}

Most strikingly, in contrast to the high-$T_{\mathrm{c}}^{\max}$ compounds, $\Gamma_{0,B1g}(T)$ becomes constant or weakly insulating only below 5\% doping [see Fig.~\ref{fig:LSCO_Gamma}].
While the temperature dependence of $\Gamma_{0,B1g}(T)$ is metallic for all $x \ge 0.05$, the big difference in the magnitude of $\Gamma_{0,B1g}(T)$ at room temperature continues to exist for $0.05 \le x \le 0.20$ [see Fig.~\ref{fig:LSCO_Gamma}].
We show that we can trace back the metallicity of $\Gamma_{0,B1g}(T)$  to an extra contribution specific to LSCO as suggested by Dumm \textit{et al.} \cite{Dumm:2002} and Venturini \textit{et al.} \cite{Venturini2002_P66_060502}.
In Fig.~\ref{fig:LSCO_UD_series} we compare data in the range $0 \le x \le 0.17$.
For comparison, samples LSCO-02 and LSCO-10 from previous measurements \cite{Tassini2005_PRL95_117002} are included.
Up to 12\% doping we observe extra peaks in the energy range below 200\,cm$^{-1}$ developing at low temperature, first in \Boneg symmetry [Fig.~\ref{fig:LSCO_UD_series} (e),(f)], then, at and below 5\%, in \Btwog symmetry [Fig.~\ref{fig:LSCO_UD_series} (h)-(j)].
Note the non-monotonic doping dependence of the low-energy peaks, which most likely is the result of slightly different crystallographic order in the samples.
The new features clearly change symmetry close to 5\% doping.
The transition is continuous in that remainders of  the additional response can be observed in \Boneg symmetry also for $x \le 0.05$.
The peak energy of the low-energy structures decreases proportional to temperature at all doping levels while the peak height increases.

%%%%%%%%%%%%%%%%%%%%%%%%%%%%%%%%%%%%%%%%%%%%%%%%%%%%%%%%%%%%%%%%%%%
%%%%%%%%%%%%%%%%%%%%%%%%%%%%%%%%%%%%%%%%%%%%%%%%%%%%%%%%%%%%%%%%%%%%%%%

\subsection{Tl$_{2}$Ba$_2$Cu$_2$O$_{6+\delta}$}
\label{sec:32}

\begin{figure*}
  \centering
  \includegraphics{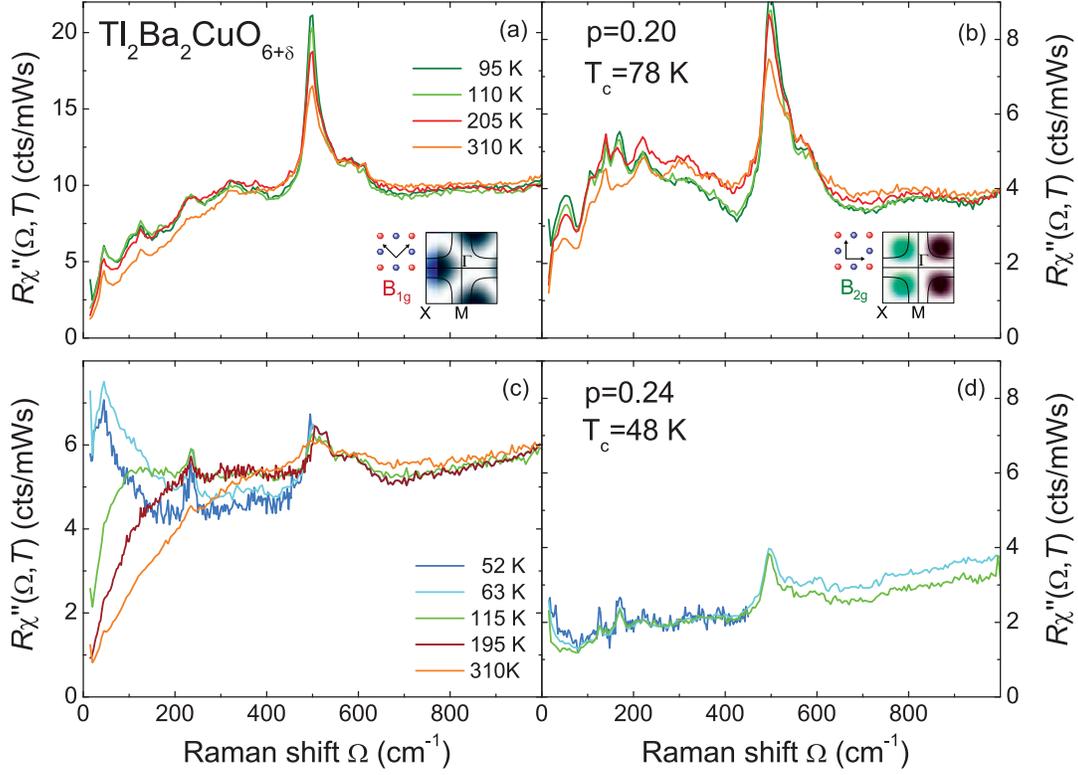}
  \caption[]{
  Raman response of Tl2201 in \Boneg and \Btwog symmetries at doping levels as indicated. Due to the small sample size and surface contamination the quality of the spectra at low energies is not as good as desirable, and there are strong contributions from the laser line. (c) For $p=0.24$ there is a strong temperature dependence of the spectra in \Boneg symmetry with a quasiparticle maximum at low temperature. (a) At $p=0.20$ the temperature dependence in \Boneg symmetry is absent.
  } \label{fig:Tl2201_spectra}
\end{figure*}

Tl2201 can be used for experiments on the overdoped side in the range $0.16 \le p \le 0.27$.
Here, we look at two samples above and below the crossover point found in Bi2212 at approximately 21\% \cite{Venturini2002_PRL89_107003}.
As shown in Fig.~\ref{fig:Tl2201_spectra}, the \Boneg spectra exhibit qualitatively different temperature dependences.
At $p=0.20$ the response changes only weakly with temperature.
In contrast, the initial slope increases rapidly at low temperatures for $p=0.24$.
In addition, a Fermi liquid maximum with an energy close to $\Gamma_{0,B1g}(T)$ emerges out of the flat continuum typical for most of the cuprates.
We note that the spectral shapes are very similar to those in Bi2212 and also LSCO at comparable doping levels (see also Ref.~\cite{Prestel2010}).
%%%%%%%%%%%%%%%%%%%%%%%%%%%%%%%%%%%%%%%%%%%%%%%%%%%%%%%%%%%%%%%%%%%
%%%%%%%%%%%%%%%%%%%%%%%%%%%%%%%%%%%%%%%%%%%%%%%%%%%%%%%%%%%%%%%%%%%%%%%%

\section{Discussion}
\label{sec:4}
\subsection{Spin excitations}
\label{sec:41}
The measurement of light scattering cross sections at high energy transfers requires a careful analysis since corrections for the response of the setup are large (up to factors of 20 for 8000\,cm$^{-1}$ shift) and the spectra may be contaminated by spurious photons in the laboratory and fluorescence of adsorbates.
Luminescence in the sample may also play a crucial role (see Fig.~\ref{fig:Bi2212_aging}) and cannot easily be controlled.
However, important information from spin and charge excitations can be obtained which is not directly accessibly by other methods.
While most of the charge excitations are observable by optical spectroscopy, the observation of spin excitations above 1200-1600\,cm$^{-1}$ remains challenging.

\begin{figure}
  \centering
  \includegraphics{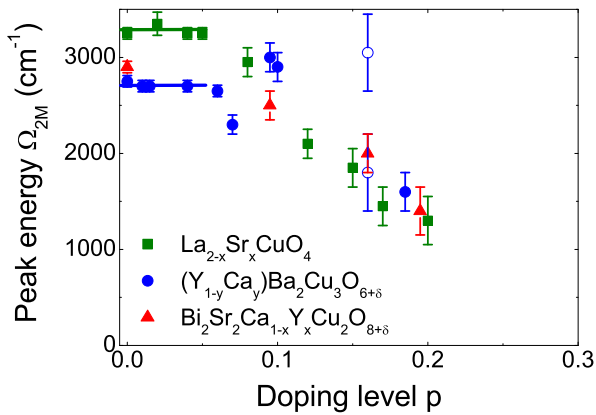}
  \caption[]{
  Peak energies of the two-magnon excitations $\Omega_{2M}$ versus doping level found for Y123 \cite{Tassini2008_PhdThesis,Opel2000_Dissertation} (dots), Bi2212 \cite{Opel2000_Dissertation} (triangles) and LSCO (squares). In the range $0.00\leq p\leq 0.05$ the two-magnon peak is essentially constant. In Y123 we found an unusually high two-magnon energy at $p=0.10$ which we trace back to strain induced by the surface treatment. At $p=0.16$ there is even a double structure as indicated by two open circles. Only at fully oxygenated and ordered Y123 ($p=0.18$) $\Omega_{2M}$ is in the expected range and well defined. The horizontal lines at low doping should emphasize that there is little variation of $\Omega_{2M}$ below the onset of superconductivity at $p_{\rm sc1} \approx 0.05$.
  } \label{fig:2magnon}
\end{figure}

Our experiments at large energy transfers focussed mainly on charge excitations and will be discussed in Ref.~\cite{Prestel2010}.
Here we make a few remarks on two-magnon excitations.
Only at zero doping are there several well defined structures in the spectra at all symmetries [Fig.~\ref{fig:LCO_puresym}].
The strong peak in \Boneg symmetry at 3300\,cm$^{-1}$ originates from nearest-neighbor spin flips.
The excitations at \Aoneg, \Atwog, and \Btwog symmetry come from spin-spin interactions with coupling beyond nearest-neighbor sites, from cyclic exchange \cite{Delannoy2005_PRB72_115114,Vernay2007_PRB75_020403}, and from coupling to the chirality operator \cite{Sulewski1991_PRL67_3864,Gozar2005__755}.
The most recent review can be found in Ref.~\cite{Gozar2005__755}.
With increasing sample quality two-magnon excitations were found in the Raman spectra up to the overdoped range (see Fig.~\ref{fig:2magnon}).
In principle, two-magnon scattering should be observable as long as clusters with correlation lengths of 3 or 4 lattice constants exist but, apparently, defects broaden the peaks sufficiently to make them to disappear into the continuum.
Fleury \textit{et al.} report two-magnon scattering for Y123 up to optimal doping \cite{Lyons1988_PRL60_732} while in high-quality Bi2212 samples, observation is possible up to $p \approx 0.20$ \cite{Rubhausen1999_PRL82_5349}.
In LSCO we find indications up to 25\%.
The maxima we observe are weak and discernable only at low temperature, but surprisingly narrow with a FWHM of only 500\,cm$^{-1}$ or less at energies above 1200\,cm$^{-1}$.
Although the results at the highest dopings should be considered with care, they have a clear correspondence to incommensurate spin fluctuations observed in neutron experiments \cite{Wakimoto2004_PRL92_217004}.
The energies of the two-magnon maxima $\Omega_{2M} \approx 2.8\,J$ are compiled in Fig.~\ref{fig:2magnon}.
As already observed by Tassini and coworkers \cite{Tassini2008_PRB78_020511}, $\Omega_{2M}$ is essentially constant below the onset of superconductivity at $p_{\rm sc1} \approx 0.05$ then jumps by some 200\,cm$^{-1}$ and decreases continuously with further doping.
This could indicate that some excitation starts to couple to the spin waves above $p_{\rm sc1}$ and renormalizes their energy.
It is tempting to conclude that the orientation change of the stripes \cite{Fujita2002_PRB65_064505} which occurs at the same doping and superconductivity itself are related to this renormalization.

The results presented here [Fig.~\ref{fig:LSCO_long}] are in disagreement with those obtained by Sugai and coworkers \cite{Sugai2003_PRB68_184504}.
The main discrepancy is their spectral shape above 2000-4000\,cm$^{-1}$.
In contrast to our results and those by Sulewski and coworkers \cite{Sulewski1991_PRL67_3864} and Reznik \textit{et al.} \cite{Reznik1993_PRB48_7624} the spectra shown in Fig.~9\,(b) of Ref.~\cite{Sugai2003_PRB68_184504} decrease more rapidly.
This leads to doping dependent peaks since the spectra lose intensity with doping [Fig.~\ref{fig:LSCO_long}].
Sugai \textit{et al.} interpret the peaks at 2000-4000\,cm$^{-1}$  at doping levels above $x=0.10$ in terms of magnons.
We do not believe that this peak is related to spin excitations in the range $0.10 \le x \le 0.25$, but rather conclude that those maxima are artifacts of the sensitivity correction.
According to our results, the lowest two-magnon energy observable at $x = 0.25$ is 1200\,cm$^{-1}$ (corresponding to $J \approx 450$\,cm$^{-1}$) rather than 500\,cm$^{-1}$ ($J \approx 200$\,cm$^{-1}$) as in Ref.~\cite{Sugai2003_PRB68_184504}.
We show in the second contribution \cite{Prestel2010} that the temperature dependent maximum in the \Boneg spectra in the range 200 to 600\,cm$^{-1}$ can be explained directly in terms of particle-hole excitations.
In materials with sufficient purity these maxima crop up for doping levels $x=p > 0.20$.
This can be masked by inhomogeneities in LSCO in the doping range $0.20 \le x \le 0.30$ for various reasons.

%%%%%%%%%%%%%%%%%%%%%%%%%%%%%%%%%%%%%%%%%%%%%%%%%%%%%%%%%%%%%%%%%%%
%%%%%%%%%%%%%%%%%%%%%%%%%%%%%%%%%%%%%%%%%%%%%%%%%%%%%%%%%%%%%%%%%%%%%%%%

\subsection{Charge dynamics}
\label{sec:42}
LSCO and Tl2201 are laboratories for studying the charge dynamics on the overdoped side.
In the case of LSCO, oxygen deficiencies can completely wipe out the intrinsic behavior \cite{Nakamae2003_PRB68_100502}.
As shown in Fig.~\ref{fig:LSCO_anneal}, the Raman spectra of overdoped samples are changed sufficiently by defects to make them look like those at lower doping levels.
However, we do not believe it is the doping level but rather the impurity concentration which is at the origin of the changes since the spectra at all doping levels are influenced in the same way and the defect-induced phonon modes almost completely disappear after annealing (Fig.~\ref{fig:LSCO_anneal}).
In ordered samples, the electronic response becomes more strongly temperature dependent and the relaxation rates $\Gamma_0$ decrease, as can be seen directly from the spectra (Fig.~\ref{fig:LSCO_anneal}) and also from the Raman resistivities (Fig.~\ref{fig:LSCO_Gamma}).
In addition, the ratio $I_{\mathrm{B1g}}/I_{\mathrm{B2g}}$ (Fig.~\ref{fig:LSCO_ratio}) is higher in ordered samples.
Intuitively one would argue that long range order enhances the density of states related to the van Hove singularity and, consequently, the \Boneg Raman intensity projecting out this part of the Brillouin zone.

For lower dopings, $0.16 < x < 0.25$, the spectra become flatter in an energy range as large as 1000\,cm$^{-1}$ in a way similar to that found by Venturini \textit{et al.} \cite{Venturini2002_PRL89_107003} for Bi2212.
In a comparable doping range, flat spectra are also found in Y123 \cite{Opel2000_PRB61_9752}.
This happens at doping levels in which quasiparticle peaks are still observed by photoemission for the entire Fermi surface \cite{Damascelli2003_RMP75_473}.
A similar transition is now also found in Tl2201.
The response at $p=0.24$ in the \Boneg channel has a quasiparticle maximum and a strong temperature dependence [Fig.~\ref{fig:Tl2201_spectra}~(c)] comparable to that of the resistivity.
At $p=0.20$, however, both the peak and the variation with $T$ are gone [Fig.~\ref{fig:Tl2201_spectra}~(a)].
We conclude that the ``unconventional metal-insulator transition'' (nomenclature of Ref.~\cite{Venturini2002_PRL89_107003}) is a common feature of a large number of hole-doped cuprates and probably generic.
According to the results in Bi2212 the transition is relatively abrupt. Andergassen \textit{et al.} explain the crossover in terms of charge-ordering fluctuations \cite{Andergassen2001_PRL87_056401} which lead to short-range nematic order at approximately 19\% doping while the mean field transition at 21\% is suppressed by fluctuations.

On the way towards the parent antiferromagnet there are two more crossover points, one close to optimal doping, $x=p \approx 0.16$, and the second one at the onset point of superconductivity, $x=p \approx 0.05$.
At $p \approx 0.16$, $\partial\Gamma_{0,B1g}(T)/\partial T$ changes sign from positive to negative in Bi2212 \cite{Venturini2002_PRL89_107003}.
This implies that the relaxation rate increases towards lower temperature. Hence, weakly insulating behavior prevails on the anti-nodal parts of the Fermi surface while the nodal parts remain metallic to much lower temperatures.
This dichotomy of the scattering rates on different parts of the Fermi surface around optimal doping is also found in ARPES experiments \cite{Chang2008_PRB78_205103}.
In conventional transport, insulating characteristics are only found well below \Tc up to optimal doping when superconductivity is suppressed by large magnetic fields \cite{Ando1995_PRL75_4662,Ando1996_PRL77_2065,Ando1997_PRB56_8530,Boebinger1996_PRL77_5417}.
$\Gamma_{0,B2g}(T)$ remains unchanged in the whole doping range down to $p \approx 0.05$ (and actually even for $p < 0.05$ but for other reasons).
Also the optical data vary slowly as a function of doping \cite{Basov2005_RoMP77_721}.
The main change at low doping is the appearance of the pseudogap \cite{Puchkov1996_JPCM8_10049} below $0.16$.
The modulation is stronger in IR than in \Btwog Raman, since IR develops more sensitivity around the nodes than the nodal Raman spectra.
Both methods demonstrate that the coherence of the nodal quasiparticles remains intact down to at least $0.05$.
In  LSCO, $\partial\Gamma_{0,B1g}(T)/\partial T$ remains positive down to at least 5\% doping and only the ratio $I_{\mathrm{B1g}}/I_{\mathrm{B2g}}$ exhibits a qualitative change at $x=p \approx 0.16$ (cf. Figs.~\ref{fig:LSCO_Gamma} and \ref{fig:LSCO_ratio}).
The reason for the differences between LSCO on one hand and Bi2212 and Y123 on the other hand can be identified in the \Boneg spectra in the doping range $0.08 \leq p=x \leq 0.12$, where an extra response develops (Fig.~\ref{fig:LSCO_UD_series}) which is either completely missing in the other compounds or not sufficiently narrow to be recognized as a separate peak.
Since this feature piles up towards low temperature, the initial slope increases strongly and makes the characteristics to appear metallic.
How do we know that these spectral structures are not coming from simple particle-hole excitations?

The answer is twofold.
First, there is experimental evidence for the existence of incipient charge and spin order \cite{Kivelson2003_RMP75_1201,Cheong1991_PRL67_1791,Tranquada1995_N375_561,Bianconi1996_PRL76_3412,Tranquada2005__} and there is a microscopic model for the response of fluctuations of incipient charge order which describes the spectral shape, symmetry and temperature dependences quantitatively \cite{Caprara2005_PRL95_117004}.
In addition, in agreement with neutron scattering results \cite{Fujita2002_PRB65_064505}, we find the response from fluctuations to flip symmetry at $x_{\mathrm{c}} \approx 0.05$ from \Boneg ($x>x_{\mathrm{c}}$) to \Btwog ($x\leq x_{\mathrm{c}}$) while maintaining all other properties (see next section). $x_{\mathrm{c}}$ is very close if not equal to $p_{\rm sc1}$.
Since there is no sum rule in electronic Raman scattering, the response just pops up and is superposed on the particle-hole continuum.
Hence, the initial slope does not track the relaxation rate of ordinary carriers but rather the characteristic energy of the fluctuation peaks.
This explains why $\Gamma_{0,B2g}(T)$ remains metallic for very low doping levels down to the lowest temperatures in spite of a non-metallic resistivity.

\subsection{Ordering fluctuations}
\label{sec:43}
The additional response from fluctuations due to incipient density-wave formation is found for LSCO in the doping range $0.02 \leq x \leq 0.12$  but not for the undoped sample (Fig.~\ref{fig:LSCO_UD_series}).
Earlier, indications at 2 and 10\% have been reported \cite{Venturini2002_P66_060502,Tassini2005_PRL95_117002}.
Caprara and coworkers explained the spectra quantitatively in terms of Aslamazov-Larkin (AL) type diagrams where two fluctuations with opposite momenta are excited simultaneously \cite{Caprara2005_PRL95_117004}.
In Y123, charge-ordering fluctuations were observed only in non-superconducting samples at low doping \cite{Tassini2008_PRB78_020511}, where only indirect support for ordering is found \cite{J'anossy2003_PRL91_177001}.
Recently, an ordering instability in superconducting samples was suggested to be at the origin of the \Tc depression \cite{Liang2006_PRB73_180505} and of the new frequency in the SdH oscillations found close to 1/8 doping \cite{LeBoeuf2007_N450_533,Pfleiderer2007_N450_492}.
Nevertheless, ordering in the compounds with high $T_c^{\max}$ is much farther from being static than in LSCO or ${\rm La_{2-x}Ba_{x}CuO_{4}}$ \cite{Tranquada2007__,Li2007_PRL99_067001}, and this may explain why no Raman response from fluctuations exists or can be resolved in superconducting  Y123 or Bi2212.

In LSCO, the low energy AL-modes due to a density wave instability are present below (\Btwog) and above (\Boneg) the doping level $p_{\rm sc1}$ at which superconductivity appears.
Even if the fluctuation response flips symmetry at $p_{\rm sc1}$, the variation with temperature and doping remains essentially the same on either side of $p_{\rm sc1}$, i.~e. the response is independent of the scattering channel in which it appears.
The selection rules, which are better defined the more the system is ordered (compare Fig.~\ref{fig:LSCO_UD_series}, and the results on Y123 in Ref.~\cite{Tassini2008_PRB78_020511}), demonstrate that the underlying order is nematic \cite{Kivelson2003_RMP75_1201,Kivelson2006_NP5_343,Fradkin2009__} which means preferential but short ranged.

\begin{figure*}
  \centering
  \includegraphics{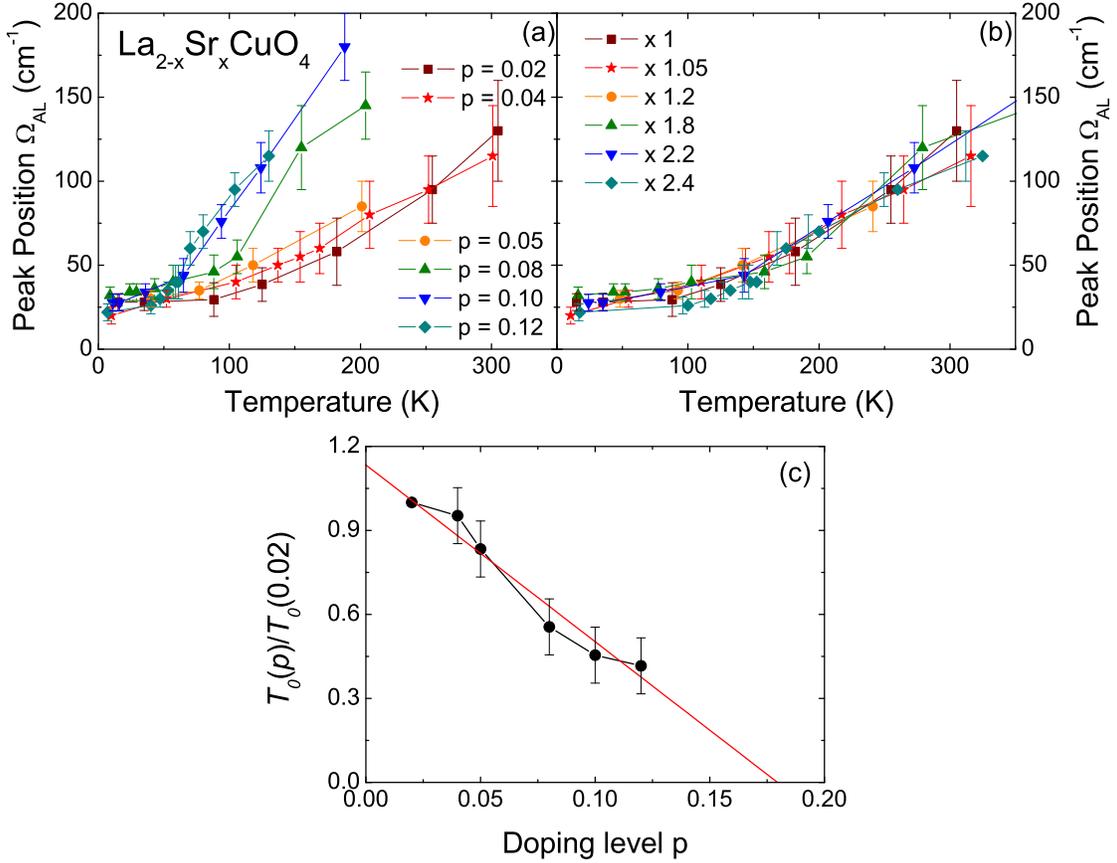}
  \caption[]{
  Peak energies of the fluctuation maxima $\Omega_{AL}$ as a function of temperature for LSCO at doping levels as indicated. (a) $\Omega_{AL}$ as derived directly from the spectra, (b) peak energies at $p>0.02$ scaled to those at $p=0.02$. The errors are due to another feature occuring at low temperature and the broadening at high temperatures. (c) Doping dependence of the scaling temperatures $T_0$. The error bars indicate the uncertainty of fitting the curves at higher doping levels on top of the curve at $p=0.02$ in (b).}
  \label{fig:LSCO_ALpeak}
\end{figure*}

In Fig.~\ref{fig:LSCO_ALpeak}~(a) we plot the peak energies of the fluctuation maxima $\Omega_{AL}$ as a function of temperature for various doping levels.
At high doping, the response exists in smaller temperature ranges than at low doping.
Similarly to Sugai and coworkers \cite{Sugai2006_PRL96_137003}, we do not observe a coexistence of the fluctuation and the superconducting coherence peaks [compare Fig.~\ref{fig:LSCO_UD_series} panels (e)-(g)].
The characteristic energies of the modes $\Omega_{AL}$ decrease with decreasing temperature and saturate below some temperature $T_0(p)$ in the range $20-30$\,cm$^{-1}$ for all doping levels.
The crossover temperature $T_0(p)$ marks the establishment of partial order.

The existence of a light scattering response well above $T_0(p)$ distinguishes the AL-modes from the usual amplitude and phase modes \cite{Lee1974_SSC14_703,Gruner1994__} in density wave systems.
The amplitude and phase modes originate from a coupling of the condensate's phase and density fluctuations to the lattice distortion, and exhibit order parameterlike behavior with the opposite temperature dependence, i.e. the energy increases with decreasing temperature.
We have shown recently for the 2D charge density wave (CDW) system DyTe$_3$ that the amplitude mode becomes very broad with increasing temperature and disappears over a range of 1\,K at the CDW transition \cite{Lavagnini2010_PRB81_081101}.
(Similar results were obtained for K$_2$SeO$_4$ \cite{Lee1988_PRB37_6442} and other systems \cite{Gruner1994__} a while ago.)
The phase mode has odd symmetry and cannot be observed by Raman scattering in systems with inversion symmetry.
These considerations lead us to disagree with the interpretation of the low-energy response in terms of collective modes of the condensate (phase and amplitude mode) as proposed by Sugai and coworkers \cite{Sugai2006_PRL96_137003}.
Actually, we found narrow modes only in disordered samples such as that at $x=0.08$ [Fig.~\ref{fig:LSCO_UD_series}~(e)] \cite{Fujita_008}.

In contrast to the phase and amplitude modes, the AL collective modes exist only above the transition and are expected to vanish if long-range order is being established.
The finite intensity and energy we find at low temperature indicate the persistence of fluctuations in the limit $T \rightarrow 0$, hence a transition from thermal to quantum fluctuations as is typical for a system close to a quantum critical point (QCP).
The location of the QCP can be determined through a scaling argument.
If we normalize the temperature scale in Fig.~\ref{fig:LSCO_ALpeak}~(a) to the crossover temperature of every doping level $T_0(p)$, the temperature dependences of the AL peaks at all doping levels collapse on top of each other as shown in Fig.~\ref{fig:LSCO_ALpeak}~(b).
Since $T_0(p)$ is ill defined, we instead use a dimensionless scaling factor which can be thought of as $f_{\rm QCP}=T_0/T_0(p)$.
For convenience, we stretch the temperature scale for samples at higher doping to match the results with those at $p=0.02$, and therefore the scaling factor is $f_{\rm QCP}=T_0(0.02)/T_0(p)$.
The scaling factors compiled in Fig.~\ref{fig:LSCO_ALpeak}~(c) are plotted as $1/f_{\rm QCP}$ and the figure clearly establishes a doping level where $1/f_{\rm QCP} \propto T_0(p)$ approaches zero.
By the data a linear extrapolation is suggested and this predicts a quantum critical point at $p_{\rm QCP}=0.18\pm0.01$, close the level predicted by Andergassen and collaborators \cite{Andergassen2001_PRL87_056401} and seen in various other experiments \cite{Tallon2001_PC349_53,Billinge2003_IJMPB17_3640}.
%%%%%%%%%%%%%%%%%%%%%%%%%%%%%%%%%%%%%%%%%%%%%%%%%%%%%%%%%%%%%%%%%%%
%%%%%%%%%%%%%%%%%%%%%%%%%%%%%%%%%%%%%%%%%%%%%%%%%%%%%%%%%%%%%%%%%%%%%%%%
\subsection{Phase diagram}
\label{sec:44}
The QCP is close to the doping level below which the superfluid density and coherence seem to decrease.
Indications have been found in many experiments and have been summarized in, e.g., Refs.~\cite{Tallon2001_PC349_53,Loram2004_PRB69_060502}.
In our experiments, we derive the location of the QCP from the carrier dynamics.
From a theoretical point of view, the response does not enter in lowest order but through a higher order process, the exchange of two fluctuations.
This may, in an intuitive way, explain why the \Boneg Raman response from the anti-nodal part of the Brillouin zone but also other two-particle susceptibilities such as the NMR relaxation rate \cite{Alloul1989_PRL63_1700,Billinge2003_IJMPB17_3640,Gutmann2000__} get depressed or incoherent much earlier than the single-particle response \cite{Damascelli2003_RMP75_473}.
We suggest that the suppression of the \Boneg coherence peaks in the superconducting state \cite{Opel2000_PRB61_9752,Nemetschek1997_PRL78_4837,Munnikes2007__} can also be traced back to fluctuations of incipient charge order.
This, of course, implies that the fluctuations are also present in Y123 or Bi2212 with high $T_c^{\max}$ even if they do not manifest themselves in an AL-type response at $p>0.05$.
However, fluctuations are found in various experiments in the static limit or, in most of the cases, in the inelastic channel \cite{Fujita2002_PRB65_064505,Kivelson2003_RMP75_1201,Cheong1991_PRL67_1791,%
Tranquada1995_N375_561,Bianconi1996_PRL76_3412,Tranquada2005__,Billinge2003_IJMPB17_3640}.

The phenomenology fits into the scenario proposed by Andergassen and coworkers \cite{Andergassen2001_PRL87_056401} on the basis of a Hubbard-Holstein model.
They propose a phase diagram which shows a mean field transition at $p \approx 0.21$ and a QCP at $p \approx 0.19$.
At present we do not have the resolution in doping to pinpoint our crossover lines sufficiently precisely. However, it appears that $\Gamma_{0,B1g}(T)$ is renormalized already at the mean field transition while the AL peaks appear only at $p \approx 0.19$.

\begin{figure}
  \centering
  \includegraphics{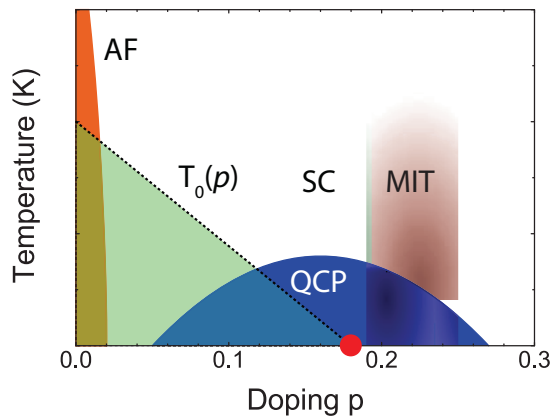}
  \caption[]{
  Schematic phase diagram of hole doped cuprates as determined from Raman experiments. AF denotes the antiferromagnetically ordered region and SC the region where superconductivity appears. The QCP is the point where the crossover temperature $T_0(p)$ extrapolates to zero. The gray shaded area indicates the doping range where the ``unconventional metal insulator transition'' (MIT) occurs.

  } \label{fig:phasediagram}
\end{figure}

Fig.~\ref{fig:phasediagram} shows the generic phase diagram of the cuprates including all transitions observed in the carrier dynamics by Raman scattering.
A preliminary version has been published in Ref.~\cite{Tassini2008_PRB78_020511}.
The QCP is the point at which the ordering temperature $T_0(p)$ extrapolates to zero.
The vertical line at $p=0.21$ indicates the ``unconventional metal-insulator transition'' \cite{Venturini2002_PRL89_107003} where the dichotomy between the nodal and anti-nodal parts of the Fermi surface appears in Bi2212.
The respective lines for LSCO and for Tl2201 are in the range $0.20 < x=p < 0.25$ and that of Y123 is at $p > 0.19$.
Hence the region around $p=0.21$ marks the border of coherence beyond which the ground state is Fermi liquid like.

In Y123 we find a very pronounced transition between super\-conduct\-ing and non-super\-conduct\-ing samples at $p=0.05$ \cite{Tassini2008_PRB78_020511}.
The two-magnon energy and width, the electronic continuum, the AL response, and the oxygen buckling mode at 340\,cm$^{-1}$ all change abruptly (on the scale of the limited doping resolution we presently have), suggesting a first-order transition.
This implies the coexistence of competing phases and favors a phase separation scenario \cite{Fradkin2009__,Grilli1991_PRL67_259,Carlson2002__} at the lower onset point of superconductivity $p_{\rm sc1}$.
Below $p_{\rm sc1}$ there is AF order which is  suppressed in a continuous fashion.
This can be derived directly from the essentially constant two-magnon energy  $\Omega_{2M}$ (Fig.~\ref{fig:2magnon}) in the range $0.00 \le p \le 0.05$.
As soon as carriers are introduced in this doping range they cluster and apparently form stripes and nematic order \cite{Fradkin2009__}.
Above $p_{\rm sc1}$ the order is oriented differently.
Only in LSCO do we find clear indications for a diagonal pattern.
In Y123 and perhaps also in Bi2212 there are only indirect indications such as increased magnon and phonon damping, all in \Boneg symmetry, which would be compatible with diagonal or checkerboard-type order.
Given the assumption that long-range density wave order and superconductivity are mutually exclusive, the suppressed $T_c$ of LSCO would just be a result of a closer proximity to the CDW ordered state.

The last pivotal point in the phase diagram is on the overdoped side where superconductivity disappears, $p_{\rm sc2}$.
Wakimoto \textit{et al.} \cite{Wakimoto2004_PRL92_217004} found the incommensurate spin order also to disappear there.
If one examines the \Boneg gap maxima on the overdoped side of the phase diagram \cite{Hackl2005_45_227,Venturini2002_PRL89_107003,Munnikes2007__,Kendziora1995_PRB52_9867,Sugai2000_PRL85_1112,LeTacon2006_NP2_537}, it becomes clear that the intensity increases beyond $p=0.21$, which can hardly be traced back to coherence.
Actually, the increase is so fast that a divergence point may be anticipated.
To explore this possibility Munnikes and coworkers \cite{Munnikes2007__} plot the inverse of the \Boneg intensity, finding a divergence point at $p=0.26\pm0.01$, which is close to $p_{\rm sc2}$.
If this were the case the \Boneg Raman response in the superconducting state would not just reflect the maximum gap but rather a superposition of the pair-breaking maximum and a mode related to a symmetry broken at $p_{\rm sc2}$.

%%%%%%%%%%%%%%%%%%%%%%%%%%%%%%%%%%%%%%%%%%%%%%%%%%%%%%%%%%%%%%%%%%%
%%%%%%%%%%%%%%%%%%%%%%%%%%%%%%%%%%%%%%%%%%%%%%%%%%%%%%%%%%%%%%%%%%%%%%%%
\section{Conclusions}
\label{sec:5}
In this publication we focussed on generic properties of hole-doped cuprates as seen by Raman scattering.

At energies above 1000\,cm$^{-1}$ the Raman response generally increases with energy.
However, the electronic continua lose intensity with doping in all symmetries and all compounds.
The synopsis of various compounds and the analysis of extrinsic effects allow us to conclude that the increase of the spectra towards high energies is an intrinsic property which has its origin in either the strong correlations in the cuprates or in the coupling between electrons and high-energy fluctuations from spin and/or charge ordering.
Since the selection rules for spin excitations are still effective and the two-magnon peaks can be resolved up to $x=0.20$, the clusters of spins must be substantially larger than the lattice constant, i.e. the correlation of just two nearest neighbor spins can be excluded for doping levels below 20\%.

The low energy continuum in \Boneg symmetry experiences a transition from spin-dominated close to zero doping to charge-dominated above $x \approx 0.15$.
Here, around optimal doping, we find changes in the carrier dynamics in many observables.
The ratio of the \Boneg to the \Btwog overall intensity and the transition between a metallic and an insulating temperature dependence of the \Boneg carrier dynamics are the most remarkable two.
In underdoped \LSCO we find a response which, based on its temperature and frequency dependence, can be traced back to density wave fluctuations.
The symmetry selection rules as obtained from the theoretical calculations \cite{Caprara2005_PRL95_117004} and the change from \Btwog to \Boneg symmetry upon crossing $p_{\rm sc1} \approx 0.05$ lead us to conclude that charge ordering is at the origin of this response.
Surprisingly, the change of symmetry does not noticeably affect the overall variation with temperature, which fits with the scenario of critical fluctuations around a quantum critical point at $x \approx 0.18$, conspicuously close to the ``unconventional metal-insulator transition'' found universally for all cuprates in the range $0.18 < p < 0.25$.
Apparently, coherence at the antinodal points $(\pi,0)$ is lost very rapidly below this doping.
To what extent the same fluctuations are also responsible for the pairing still cannot be answered.
However, the first-order like changes of practically all excitations at $p_{\rm sc1}$ suggest strong interactions between carriers and both spin and charge fluctuations.

\section*{Acknowledgements}
We profited immensely from discussions with S. Caprara, T.~P. Devereaux, C. Di~Castro,  M. Grilli, S. Kivelson, and J. Tranquada.
This work has been supported by the DFG via Research Unit FOR~538 (grant-nos. Ha2071/3 and Er342/1). B.M. and R.H. gratefully acknowledge support by the Bavarian Californian Technology Center (BaCaTeC).
Y.A. is supported by KAKENHI 20030004 and 19674002.
D.P., R.L., D.B., and W.H. were supported by NSERC Canada.

\end{document}